\documentclass{aa}  
\usepackage{graphicx}
\usepackage{txfonts}
\usepackage{CJKutf8}
\usepackage[colorlinks=true, linkcolor=blue, citecolor=blue, filecolor=blue, urlcolor=blue]{hyperref}
\begin{document} 
   \title{Planetesimal formation in a pressure bump induced by infall}

\author{
Haichen Zhao
\begin{CJK*}{UTF8}{gbsn}(赵海辰)\end{CJK*}
\inst{1}
\and
Tommy Chi Ho Lau
\begin{CJK*}{UTF8}{bsmi}(劉智昊)\end{CJK*}
\inst{1}
\thanks{\emph{Present address:}
Department of Astronomy and Astrophysics, University of Chicago, Chicago, IL 60657, USA}
\and
Tilman Birnstiel
\inst{1,2}
\and
Sebastian M. Stammler
\inst{1}
\and
Joanna Dr{\c a}{\.z}kowska
\inst{3}
}

   \institute{University Observatory, Faculty of Physics, Ludwig-Maximilians-Universit\"at M\"unchen, Scheinerstr. 1, D-81679 Munich, Germany\\
             \email{hzhao@usm.lmu.de}
         \and
             Exzellenzcluster ORIGINS, Boltzmannstr. 2, D-85748 Garching, Germany
        \and
        Max Planck Institute for Solar System Research, Justus-von-Liebig-Weg 3, 37077 Göttingen, Germany
             }

   \date{Received 12 November 2024; accepted 27 January 2025}

\abstract
{Infall of interstellar material is a potential non-planetary origin of pressure bumps in protoplanetary disks. While pressure bumps arising from other mechanisms have been numerically demonstrated to promote planet formation, the impact of infall-induced pressure bumps remains unexplored.}
{We aim to investigate the potential for planetesimal formation in an infall-induced pressure bump, starting with sub-micrometer-sized dust grains, and to identify the conditions most conducive to triggering this process.}
{We developed a numerical model that integrates axisymmetric infall, dust drift, and dust coagulation, along with planetesimal formation via streaming instability. Our parameter space includes gas viscosity, dust fragmentation velocity, initial disk mass, characteristic disk radius, infall rate and duration, as well as the location and width of the infall region.}
{An infall-induced pressure bump can trap dust from both the infalling material and the outer disk, promoting dust growth. The locally enhanced dust-to-gas ratio triggers streaming instability, forming a planetesimal belt inside the central infall location until the pressure bump is smoothed out by viscous gas diffusion. Planetesimal formation is favored by a massive, narrow streamer infalling onto a low-viscosity, low-mass, and spatially extended disk containing dust with a high fragmentation velocity. This configuration enhances the outward drift speed of dust on the inner side of the pressure bump, while also ensuring the prolonged persistence of the pressure bump. Planetesimal formation can occur even if the infalling material consists solely of gas.}
{A pressure bump induced by infall is a favorable site for dust growth and planetesimal formation, and this mechanism does not require a preexisting massive planet to create the bump.}


\maketitle

\section{Introduction} \label{sec:intro}

The formation of planetesimals, typically hundreds of kilometers in diameter, from micrometer-sized dust particles is a fundamental stage in planetary evolution according to the core accretion theory \citep{1996Icar..124...62P}. A significant challenge in this process is generally known as the "meter-size barrier", which refers to the difficulty in growing dust particles beyond one meter due to rapid radial drift and destructive collisions \citep{2008ARA&A..46...21B}. One potential solution to overcome this barrier is the presence of dust traps --- regions where the aerodynamic drag force on dust grains weakens or vanishes entirely, facilitating dust accumulation and growth. 
Pressure bumps, a typical kind of dust trap arising from a local maximum in gas pressure, have received increasing attention in recent years as a favorable site for planet formation \citep[e.g.,][]{2020A&A...638A...1M,2020A&A...642A.140G,2021ApJ...914..102C,2023MNRAS.518.3877J,2024A&A...686A..78S,2022A&A...668A.170L,2024A&A...688A..22L}.
Since dust particles drift toward higher pressure \citep{1977MNRAS.180...57W,1986Icar...67..375N,2002ApJ...581.1344T}, pressure bumps act as barriers to inward drifting dust, accumulating the dust from the outer disk and increasing dust concentration within the bump. The locally enriched dust-to-gas ratio could trigger streaming instability \citep{2005ApJ...620..459Y,2007Icar..192..588Y,2007Natur.448.1022J,2009ApJ...704L..75J,2010ApJ...722.1437B}, leading to the formation of planetesimals. The compositional and dynamical properties of planetesimals that formed through streaming instability are generally consistent with those of comets and Kuiper Belt objects \citep{2017MNRAS.469S.755B,2019NatAs...3..808N}.

Recent high-resolution interferometry observations from the Atacama Large Millimeter/submillimeter Array (ALMA) have shown that substructures are prevalent in protoplanetary disks. These substructures primarily appear as axisymmetric rings, as documented in various surveys \citep[e.g.,][]{2018ApJ...869L..41A,2018ApJ...869...17L,2021MNRAS.501.2934C}. Although these observations focus on the largest and brightest disks, comparisons between observational data of disk populations and theoretical models imply that such substructures must be common even in unresolved disks \citep{2021MNRAS.507..818T,2022A&A...661A..66Z,2024A&A...688A..81D}. A detailed study by \citet{2018ApJ...869L..46D} demonstrates that dust trapping in local pressure maxima is consistent with the rings sampled from the DSHARP survey. Kinematic studies with ALMA by \citet{2018ApJ...860L..12T,2018ApJ...868..113T} further confirm that dust rings are associated with pressure bumps in the analyzed disks. \citet{2019ApJ...884L...5S} propose that planetesimal formation through streaming instability, where the midplane dust-to-gas ratio is regulated, could account for the observed similarities in the optical depths across the DSHARP rings identified by \citet{2018ApJ...869L..46D}. Hydrodynamical simulations with self-gravity by \citet{2021AJ....161...96C} indicate that even a small pressure bump could initiate planetesimal formation via streaming instability with centimeter-sized grains. However, this process may be problematic for millimeter-sized particles \citep{2022ApJ...933L..10C}, which are most accessible to (sub)millimeter observations with ALMA.

The origin of pressure bumps, however, remains uncertain. The most widely proposed mechanism is planet-disk interaction, where a massive planet opens a gap in the disk, resulting in a pressure bump just outside the gap \citep{2006MNRAS.373.1619R,2012A&A...545A..81P,2015MNRAS.453L..73D,2017ApJ...843..127D,2018ApJ...869L..47Z}. However, among the many disks observed by direct imaging, PDS 70 remains the only system where accreting exoplanets have been unambiguously identified \citep{2018A&A...617A..44K,2019NatAs...3..749H}. Furthermore, the formation of such massive planets themselves needs to be explained independently of planet-induced pressure bumps. Therefore, non-planetary mechanisms that might induce pressure bumps should also be considered, including dead zone boundaries \citep{2015A&A...574A..68F,2016A&A...596A..81P}, zonal flows \citep{2014ApJ...796...31B}, snowlines \citep{2007ApJ...664L..55K,2011ApJ...728...20S}, photoevaporation \citep{2019MNRAS.487.3702O}, and infall of material \citep{2022ApJ...928...92K}.

Recent discoveries of infall streamers indicate that infall can occur at various evolutionary stages of protostars, including Class 0 protostars, which are embedded in dense gas and dust envelopes \citep[e.g.,][]{2012ApJ...748...16T,2018ApJ...862....8T,2020NatAs...4.1158P,2022ApJ...925...32T,2022A&A...658A..53M}; Class I protostars, where protoplanetary disks are gradually forming \citep[e.g.,][]{2014ApJ...793....1Y,2022A&A...658A.104G,2022A&A...667A..12V,2023ApJ...958...98F,2024A&A...682A..61C}; and Class II protostars, where protoplanetary disks are fully developed \citep[e.g.,][]{2012A&A...547A..84T,2020ApJ...904L...6A,2021ApJ...908L..25G,2021ApJS..257...19H,2023A&A...670L...8G}. \citet{2024ApJ...972L...9W} estimate that 20\% to 70\% of protoplanetary disks accrete more than 50\% of their material through late-stage infall. While numerous works have modeled the formation and evolution of planetary cores in pressure bumps caused by various mechanisms \citep[e.g.,][]{2020A&A...638A...1M,2020A&A...642A.140G,2021ApJ...914..102C,2023MNRAS.518.3877J,2024A&A...686A..78S,2022A&A...668A.170L,2024A&A...688A..22L}, as of yet no study has simulated planet formation specifically in infall-induced pressure bumps.

In this work, we present a numerical model for the formation of planetesimals in an axisymmetric pressure bump induced by late-stage infall onto a protoplanetary disk. Section \ref{sec:method} describes the physical models for the structure and evolution of gas and dust in the disk, the infall process, and the formation of planetesimals from the accumulated dust. Section \ref{sec:setup} outlines our simulation setup using the gas and dust evolution code \texttt{DustPy} \citep{2022ApJ...935...35S}, along with the parameter space explored. The simulation results and analysis are detailed in Section \ref{sec:results}. We provide discussions of our model and results in Section \ref{sec:dis}, and conclude the study in Section \ref{sec:conc}.

\section{Model} \label{sec:method}

Our model consists of disk evolution, infall, and planetesimal formation. We assume a 1D axisymmetric protoplanetary disk orbiting around a Solar-mass central star, where the gas undergoes viscous evolution and the dust drifts and grows.

\subsection{Disk structure and evolution} \label{ssec:disk}

\subsubsection{Gas component}

The initial gas surface density profile of the protoplanetary disk adopts the self-similar solution derived by \citet{1974MNRAS.168..603L},
\begin{equation}
    \Sigma_{\rm g,init} = \frac{M_{\rm disk}}{2\pi r_{\rm c}^2} \left(\frac{r}{r_{\rm c}} \right)^{-1}\exp \left(-\frac{r}{r_{\rm c}} \right),
    \label{eq:init}
\end{equation}
where $r$ is the distance from the central star, $M_{\rm disk}$ is the initial mass of the disk, and $r_{\rm c}$ is the characteristic radius of the disk. Assuming $M_{\rm disk} = 0.05\, M_{\odot}$ and $r_{\rm c} = 100\,\rm au$, for instance, it yields $\Sigma_{\rm g,init} \approx 700\,\rm g\, cm^{-2}$ at $r = 1\,\rm au$.

The midplane gas temperature is calculated assuming a passively irradiated disk with a constant irradiation angle of 0.05 \citep{1997ApJ...490..368C,2001ApJ...560..957D},
\begin{equation}
    T_{\rm g} \approx 221.3 \left(\frac{r}{\rm au} \right)^{-1/2}\,\rm K,
\end{equation}
from which the isothermal sound speed at the midplane $c_{\rm s} = \sqrt{k_{\rm B} T_{\rm g} / \mu m_{\rm p}}$ can be calculated with the Boltzmann constant $k_{\rm B}$, the mean molecular weight of the gas $\mu = 2.3$, and the proton mass $m_{\rm p}$.

The protoplanetary disk is assumed to be in vertical hydrostatic equilibrium. Given the pressure scale height of the gas defined as $H_{\rm g} \equiv c_{\rm s} / \Omega_{\rm K}$, where $\Omega_{\rm K}$ is the Keplerian orbital frequency, the aspect ratio of the disk is quantified by
\begin{equation}
    h = \frac{H_{\rm g}}{r} \approx 0.03 \left(\frac{r}{\rm au} \right)^{1/4}.
    \label{eq:h}
\end{equation}
The midplane gas volume density can be derived according to vertical hydrostatic equilibrium as
\begin{equation}
    \rho_{\rm g} = \frac{\Sigma_{\rm g}}{\sqrt{2\pi}H_{\rm g}}.
    \label{eq:rho_g}
\end{equation}
The midplane gas pressure is then given by the ideal gas law $P = \rho_{\rm g} c_{\rm s}^2$, from which we can obtain the midplane pressure gradient parameter
\begin{equation}
    \eta = -\frac{h^2}{2} \frac{\partial \ln{P}}{\partial \ln{r}}.
    \label{eq:eta}
\end{equation}
This quantity characterizes the degree of “sub-Keplerianity” of the gas, given by the relation $v_{\text{g},\varphi}^2 = (1 - 2\eta)v_{\rm K}^2$, where $v_{\text{g},\varphi}$ is the azimuthal velocity of the gas and $v_{\rm K}$ is the Keplerian orbital velocity. For small values of $\eta$, this expression simplifies to $v_{\text{g},\varphi} = (1 - \eta)v_{\rm K}$.

The evolution of the gas follows the viscous advection-diffusion equation \citep{1952ZNatA...7...87L,1974MNRAS.168..603L,1981ARA&A..19..137P},
\begin{equation}
    \frac{\partial \Sigma_{\rm g}}{\partial t} = \frac{3}{r} \frac{\partial}{\partial r} \left[\sqrt{r} \frac{\partial}{\partial r} \left(\nu \Sigma_{\rm g}\sqrt{r} \right) \right] + S_{\rm g,infall},
\end{equation}
with the kinematic viscosity $\nu = \alpha c_{\rm s} H_{\rm g}$ adopting the $\alpha$ prescription introduced by \citet{1973A&A....24..337S}, and the external source term $S_{\rm g,infall}$ referring to the flux of the infalling gas. Neglecting the dynamic back reaction of dust onto the gas, the gas radial velocity due to viscous accretion is given by
\begin{equation}
    v_{\text{g},r} = -\frac{3}{\Sigma_{\rm g}\sqrt{r}} \frac{\partial}{\partial r} \left(\nu \Sigma_{\rm g}\sqrt{r} \right).
    \label{eq:vgr}
\end{equation}

\subsubsection{Dust component}

The initial surface density of the dust is assumed to be proportional to that of the gas,
\begin{equation}
    \Sigma_{\rm d,init} = \epsilon_{\rm init} \Sigma_{\rm g,init},
\end{equation}
where the initial global dust-to-gas ratio $\epsilon_{\rm init}$ is set to the Solar metallicity of 0.01. The initial size distribution of the dust particles $n(a)$ follows the MRN (Mathis-Rumpl-Nordsieck) size distribution of interstellar grains introduced by \citet{1977ApJ...217..425M}, expressed as $n(a) \propto a^{-7/2}$ up to the maximum initial particle radius of 1 $\rm\mu m$. A bulk mass density of $\rho_{\rm s} = 1.67 \,\rm g\,cm^{-3}$ is assumed. The Stokes number $\text{St}_i$ of each particle mass bin $m_i$ (see Section \ref{sec:setup} for details on particle mass binning) is determined by considering both the Epstein and Stokes I drag regimes.

The dust scale height of each mass bin $H_{\text{d},i}$ is derived by solving a vertical settling-mixing equilibrium equation by \citet{1995Icar..114..237D}, calculated as
\begin{equation}
    H_{\text{d},i} = H_{\rm g} \sqrt{\frac{\delta_{\rm vert}}{\delta_{\rm vert}+\text{St}_i}},
    \label{eq:H_d}
\end{equation}
where the vertical mixing parameter $\delta_{\rm vert} = \alpha$ is assumed. Then, the midplane dust volume density of each mass bin $\rho_{\text{d},i}$ is obtained via vertical hydrostatic equilibrium.

The evolution of the dust surface density $\Sigma_{\text{d},i}$ involves dust transport, dust growth, and external sources from infall. Dust transport is described by the advection-diffusion equation \citep{1988MNRAS.235..365C,2010A&A...513A..79B},
\begin{equation}
    \frac{\partial \Sigma_{\text{d},i}}{\partial t} + \frac{1}{r} \frac{\partial}{\partial r} \left[r \Sigma_{\text{d},i} v_{\text{d},i} - r D_{\text{d},i} \Sigma_{\rm g} \frac{\partial}{\partial r} \left(\frac{\Sigma_{\text{d},i}}{\Sigma_{\rm g}} \right) \right] = S_{\text{d,infall},i}
\end{equation}
where $D_{\text{d},i} = \delta_{\rm rad} c_{\rm s} H_{\rm g}/(1+\text{St}_i^2)$ is the dust diffusivity \citep{2007Icar..192..588Y} assuming the radial mixing parameter $\delta_{\rm rad} = \alpha$, and the external source term $S_{\text{d,infall},i}$ is the dust infall flux. The radial velocity of the dust $v_{\text{d},i}$, according to \citet{1986Icar...67..375N} and \citet{2002ApJ...581.1344T}, is given by
\begin{equation}
    v_{\text{d},i} \simeq -\frac{2 \eta v_{\rm K}}{\text{St}_i + \text{St}_{i}^{-1}}.
    \label{eq:drift}
\end{equation}
Dust growth by coagulation is calculated by solving the Smoluchowski equation \citep{1916ZPhy...17..557S}.

The radial drift speed of the dust rises as the particle sizes increase within $\text{St} < 1$, thus setting a limit for dust particle sizes as the drift rate exceeds the coagulation rate. Another limit is fragmentation, which occurs when the impact velocity between two dust grains exceeds a critical fragmentation velocity $v_{\rm frag}$. Laboratory experiments have measured $v_{\rm frag}$ around $100 \,\rm cm\,s^{-1}$ for silicate grains \citep{2008ARA&A..46...21B,2010A&A...513A..56G}. Water-ice particles have a significantly higher $v_{\rm frag}$ of over $1000 \,\rm cm\,s^{-1}$ due to their approximately tenfold higher surface energies, as confirmed experimentally by \citet{2015ApJ...798...34G}. The disk population study by \citet{2024A&A...688A..81D} suggests a range of $v_{\rm frag} \geq 500 \,\rm cm\,s^{-1}$. In this work, we adopt a relatively conservative range, limiting $v_{\rm frag}$ to values below $500 \,\rm cm\,s^{-1}$. For further details on the gas and dust evolution algorithm, readers can refer to \citet{2022ApJ...935...35S}.

\subsection{Infall model} \label{ssec:infall}

\begin{figure}
\resizebox{\hsize}{!}{\includegraphics{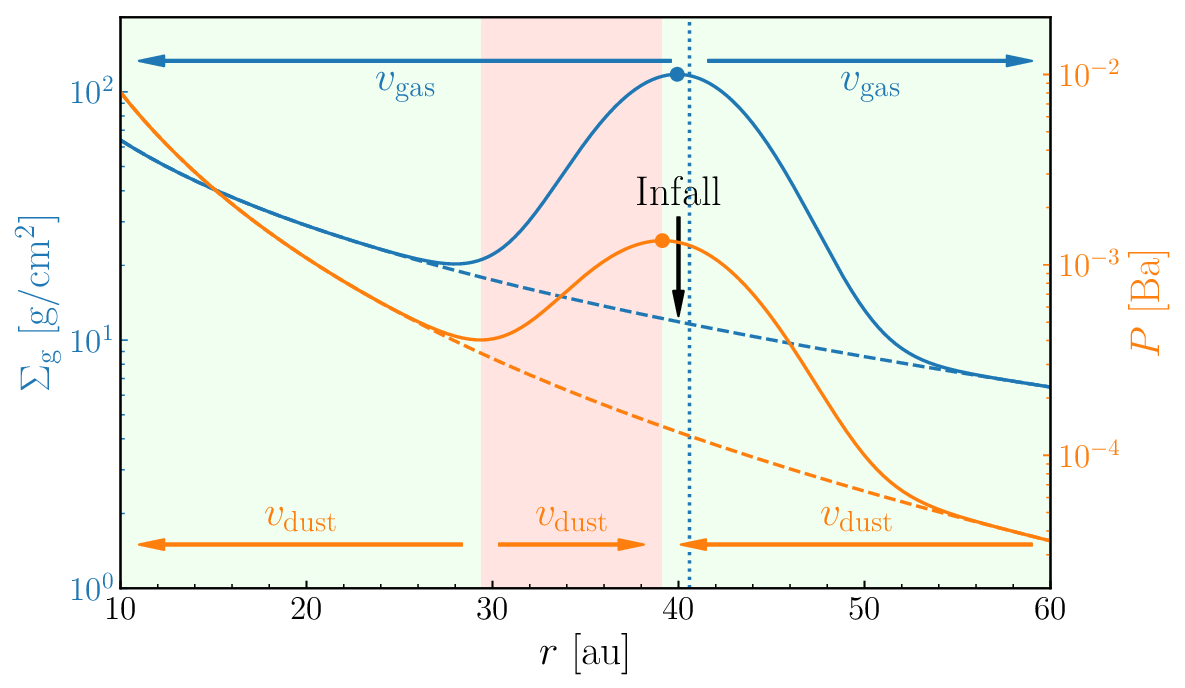}}
\caption{Illustration of the infall model, showing profiles of gas surface density (blue curves) and pressure (orange curves). The dashed curves represent the initial profiles, while the solid curves depict the profiles after infall. The horizontal blue and orange arrows indicate the directions of gas and dust radial velocities, respectively.}
\label{fig:infall}
\end{figure}

Gas and dust infall onto the protoplanetary disk over a duration of $t_{\rm infall}$, starting at $t = 0$. The infall rate is constant during this process. We define the infall rate of gas by $\dot{M}_{\rm infall} \equiv M_{\rm infall} / t_{\rm infall}$, where $M_{\rm infall}$ is the total mass of the infalling gas. The global dust-to-gas ratio of the infalling materials equals the initial condition of the disk, $\epsilon_{\rm init} = 0.01$, thus the total mass of the infalling materials consisting of gas and dust is $1.01\, M_{\rm infall}$. The initial particle sizes of the infalling dust also follow the MRN size distribution. 

The infalling materials are distributed axisymmetrically in the disk, and the flux follows a Gaussian distribution over $r$,
\begin{equation}
    S_{\rm g,infall} = \frac{\dot{M}_{\rm infall}}{A} \exp\left[-\frac{\left(r - r_{\rm infall}\right)^2}{2 \sigma_{r}^2}\right],
\end{equation}
where $r_{\rm infall}$ is the central radius of infall, assumed to be fixed over time, and $\sigma_{r}$ describes the width of the Gaussian infall. The normalizing area factor $A$ ensures the integral of $S_{\rm g,infall}$ over the disk area yields $\dot{M}_{\rm infall}$, formulated as
\begin{equation}
    A = \int_{r_{\rm min}}^{r_{\rm max}} 2\pi r \exp\left[-\frac{\left(r - r_{\rm infall}\right)^2}{2 \sigma_{r}^2}\right] \text{d}r,
\end{equation}
with the inner and outer boundaries of the disk $r_{\rm min}$ and $r_{\rm max}$.

Figure \ref{fig:infall} provides an illustration of the infall model, with the central infall radius at $r_{\rm infall} = 40$ au as indicated by the black arrow. The infall width parameter is set to $\sigma_{r} = 4$ au, and gas with a total mass of $10^4\, M_{\oplus}$ falls onto a disk with an initial mass of $M_{\rm disk} = 0.05\, M_{\odot}$ and characteristic radius $r_{\rm c} = 100\,\rm au$. A local maximum in gas surface density, marked by the blue dot, appears slightly inward of $r_{\rm infall}$. Since the sound speed $c_{\rm s}$ decreases with radial distance, a local maximum in pressure forms even closer to the star, as indicated by the orange dot. This pressure bump results in a reversal of the pressure gradient across it, distinguishing sub-Keplerian (green-shaded region, $\eta > 0$) and super-Keplerian (red-shaded region, $\eta < 0$) gas. The orange horizontal arrows illustrate the dust drift direction, which follows the pressure gradient as given by Eq. (\ref{eq:drift}). Dust particles are drawn toward the pressure peak, leading to increased dust concentration at this location. The gas radial velocity, calculated from Eq. (\ref{eq:vgr}), changes direction at the vertical dotted blue line. Over time, viscous diffusion gradually smooths the pressure bump, making the dust trap a temporary feature. Because this boundary lies beyond the local maxima in gas surface density and pressure, viscous spreading causes the pressure maximum to shift inward. Since this work neglects the back reaction of dust on gas, which generally causes the gas to flow toward lower pressure \citep{1986Icar...67..375N}, the pressure gradient does not contribute to smoothing the bump.

\subsection{Planetesimal formation} \label{ssec:plts}

As the dust accumulates and grows at the pressure bump induced by infall, the midplane dust-to-gas ratio gradually increases, and streaming instability can be triggered to transform dust into planetesimals. The prescriptions for the transformation are based on the methods in \citet{2016A&A...594A.105D} and \citet{2018A&A...620A.134S}, which considered a midplane dust-to-gas ratio,
\begin{equation}
    \epsilon_{\rm mid} = \frac{\rho_{\rm d}}{\rho_{\rm g}} = \sum\limits_{i} \frac{\rho_{\text{d},i}}{\rho_{\rm g}},
\end{equation}
exceeding unity as the condition for streaming instability \citep{2014A&A...572A..78D}. In addition, following the prescriptions in \citet{2024A&A...688A..22L}, the condition for gravitational collapse of a dense filament induced by streaming instability is determined by a Toomre-like criterion, requiring $Q_{\rm p} < 1$. The parameter $Q_{\rm p}$ is given by
\begin{equation}
    Q_{\rm p} = \sqrt{\frac{\delta}{\text{St}_{\rm avg}}} \frac{c_{\rm s} \Omega_{\rm K}}{\pi G \Sigma_{\rm d,local}} = \sqrt{\frac{\delta}{\text{St}_{\rm avg}}} \frac{Q}{\Sigma_{\rm d,local}/\Sigma_{\rm g}}
    \label{eq:Qp}
\end{equation}
as defined in \citet{2023ApJ...949...81G}. Here, $Q = c_{\rm s} \Omega_{\rm K} / (\pi G \Sigma_{\rm g})$ is the standard Toomre parameter \citep{1964ApJ...139.1217T}. The term $\delta$ represents the small-scale diffusion parameter. The density-averaged Stokes number is defined as $\text{St}_{\rm avg} = \sum\limits_{i} \text{St}_i \Sigma_{\text{d},i} / \sum\limits_{i} \Sigma_{\text{d},i}$. The local dust surface density at the dense filament, $\Sigma_{\rm d,local}$, is assumed to be ten times $\Sigma_{\rm d}$. According to the measurements in streaming instability simulations by \citet{2018ApJ...861...47S}, the value of $\delta$ ranges from $10^{-6}$ to $10^{-4}$. Since a lower $\delta$ is beneficial to planetesimal formation, we fixed $\delta$ at a conservative value of $10^{-4}$ to make planetesimal formation more challenging.

\begin{figure}
\resizebox{\hsize}{!}{\includegraphics{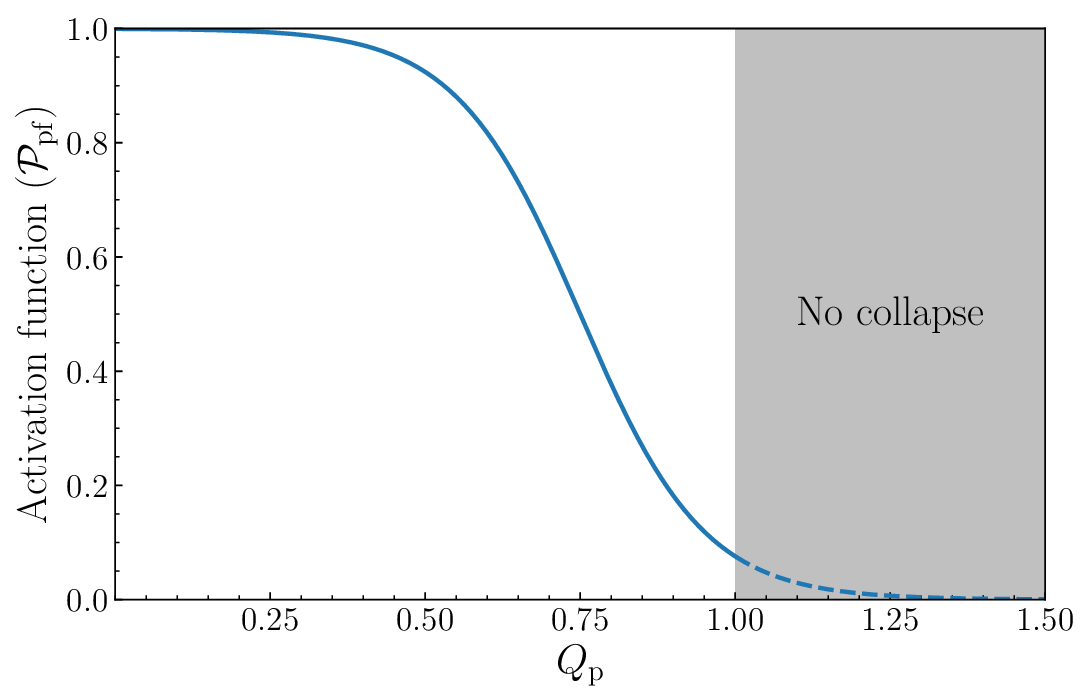}}
\caption{Smooth activation functions in terms of $Q_{\rm p}$ with smoothness parameter $n = 0.2$, centered at $Q_{\rm p} = 0.75$. Gravitational collapse occurs when $Q_{\rm p} < 1$.}
\label{fig:act}
\end{figure}

Once both conditions, $\epsilon_{\rm mid} \geq 1$ and $Q_{\rm p} < 1$, are satisfied simultaneously in a radial cell (see Section \ref{sec:setup} for details on the radial grid), the dust surface density in that cell is transformed into the surface density of planetesimals by
\begin{equation}
    \frac{\partial \Sigma_{\rm plts}}{\partial t} = - \sum\limits_{i} \left(\frac{\partial \Sigma_{\text{d},i}}{\partial t} \right)_{\rm plts},
\end{equation}
with the reduction rate of dust surface density of each dust species $i$ that contributes to planetesimal masses
\begin{equation}
    \left(\frac{\partial \Sigma_{\text{d},i}}{\partial t} \right)_{\rm plts} = - \mathcal{P}_{\rm pf} \frac{\zeta \Sigma_{\text{d},i}}{t_{\text{sett},i}} = - \mathcal{P}_{\rm pf} \zeta \Sigma_{\text{d},i} \text{St}_i \Omega_{\rm K}.
    \label{eq:plts_dt}
\end{equation}
The settling time $t_{\text{sett},i} = (\text{St}_i \Omega_{\rm K})^{-1}$ is the timescale on which streaming instability filaments form \citep{2017A&A...606A..80Y}. The planetesimal formation efficiency $\zeta$ describes the fraction of dust transformed into planetesimals per settling time, set at $10^{-3}$. The parameter $\zeta$ is not yet well understood; however, it does not impact the conditions required to trigger planetesimal formation. The term $\mathcal{P}_{\rm pf}$ is a smooth activation function designed to prevent abrupt initiation of planetesimal formation in \citet{2021MNRAS.508.5638M}, where it was defined as a hyperbolic tangent function of $\epsilon_{\rm mid}$. Unlike \citet{2021MNRAS.508.5638M}, here it is defined as a function of $Q_{\rm p}$, formulated as
\begin{equation}
    \mathcal{P}_{\rm pf} = \frac{1}{2} \left[1 - \tanh \left(\frac{Q_{\rm p} - 0.75}{n}\right) \right],
    \label{eq:act}
\end{equation}
with the smoothness parameter $n$. We set $n = 0.2$ and made the function center around $Q_{\rm p} = 0.75$ so that the value at the planetesimal formation threshold $Q_{\rm p} = 1$ is positive but rather low, as shown in Figure \ref{fig:act}. Each of our simulations eventually produces a surface density distribution of planetesimals.

In the outer regions of the disk, the relative velocity between dust grains can remain consistently below $v_{\rm frag}$, preventing fragmentation from occurring. If the pressure bump is located in this region, some dust particles will grow to $\text{St} \gg 1$ in the absence of a fragmentation limit. However, streaming instability requires dust grains that are moderately coupled to the gas \citep{2005ApJ...620..459Y,2007Natur.448.1022J,2010ApJ...722.1437B}, meaning that excessively large dust particles will not participate in streaming instability. Numerous studies have shown that when the vertically integrated dust-to-gas ratio $\epsilon$ is around the Solar metallicity ($\sim$0.01), strong clumping induced by streaming instability can only occur for dust grains with $0.01 \lesssim \text{St} \lesssim 1$ \citep{2015A&A...579A..43C,2017A&A...606A..80Y,2021ApJ...919..107L,2022arXiv221204509S,2024ARA&A..62..157B}. Therefore, all calculations from Eqs. (\ref{eq:Qp}) to (\ref{eq:act}) only consider dust species with $\text{St} < 1$. A lower limit for St is unnecessary because the mass fraction of dust species with $\text{St} \ll 0.01$ is negligible once the dust trap has formed.

\section{Simulation setup and parameter space} \label{sec:setup}

\begin{table*}
\caption{Parameters of the model, fiducial values, and ranges for random sampling.}
\label{tab:param}
\centering
\begin{tabular}{l l c c}
\hline\hline
Parameter & Description & Fiducial value & Range \\
\hline
$\alpha$ & Gas viscosity parameter & $2 \times 10^{-4}$ & [1, 4] $\times 10^{-4}$ \\
$v_{\rm frag}$ ($\rm cm\,s^{-1}$) & Dust fragmentation velocity & 500 & [100, 500] \\
$M_{\rm disk}$ ($M_{\odot}$) & Initial disk mass & 0.05 & [0.02, 0.1] \\
$r_{\rm c}$ (au) & Characteristic disk radius & 100 & [50, 200] \\
$r_{\rm infall}$ (au) & Central infall radius & 40 & [20, 80] \\
$\dot{M}_{\rm infall}$ ($M_{\oplus}\,\rm yr^{-1}$) & Gas infall rate & 0.2 & [0.1, 0.5] \\
$t_{\rm infall}$ (kyr) & Infall duration & 50 & [20, 100] \\
$\sigma$ & Relative infall width & 0.1 & [0.05, 0.2] \\
\hline
\end{tabular}
\end{table*}

Our simulations were performed using the code \texttt{DustPy}\footnote{\texttt{DustPy} v1.0.5 was used for the simulations in this work.} \citep{2022ApJ...935...35S}. \texttt{DustPy} is a \texttt{Python} package to simulate the radial evolution of gas and dust in protoplanetary disks, including viscous gas evolution, dust advection and diffusion, as well as dust growth by coagulation and fragmentation, based on the model of \citet{2010A&A...513A..79B}.

The dust particle masses are binned logarithmically from $10^{-12}$ to $10^8\,\rm g$, using a total of 141 bins. The radial grid spans from $r_{\rm min} = 1\,\rm au$ to $r_{\rm max} = 10^3\,\rm au$, with refinement around the infall region as follows: First, a logarithmic radial grid from $r_{\rm min}$ to $r_{\rm max}$ with 100 cells is created. Then, the section of the grid between $(r_{\rm infall} - 3 \sigma_{r})$ and $(r_{\rm infall} + 3 \sigma_{r})$ is replaced with a logarithmic grid consisting of 60 cells, where 99.7\% of the infall materials are dumped.

There are eight parameters considered in our model: the gas viscosity parameter $\alpha$, the dust fragmentation velocity $v_{\rm frag}$, the initial disk mass $M_{\rm disk}$, the characteristic disk radius $r_{\rm c}$, the central infall radius $r_{\rm infall}$, the gas infall rate $\dot{M}_{\rm infall}$, the infall duration $t_{\rm infall}$, and the relative width of the infall region $\sigma \equiv \sigma_{r}/r_{\rm infall}$. The value ranges of the parameters are listed in Table \ref{tab:param}. Infall begins at the start of the simulations, and all simulations terminate at $t_{\rm tot} = 500\,\rm kyr$, which is long enough compared with $t_{\rm infall}$ to determine whether planetesimals can form.

The range of $\alpha$ values aligns with those adopted in many recent simulation studies on pressure bumps \citep[e.g.,][]{2021ApJ...914..102C,2024A&A...688A..22L,2024A&A...686A..78S}. Observational estimates \citep[e.g.,][]{2018ApJ...869L..46D,2021ApJ...912..164D,2024A&A...682A..32J} indicate that $\alpha \sim 10^{-4}$ is consistent with the majority of observed disks. The value ranges of the disk parameters $M_{\rm disk}$ and $r_{\rm c}$ align with current observational measurements \citep{2020ARA&A..58..483A}. The value ranges of the infall parameters $\dot{M}_{\rm infall}$ and $t_{\rm infall}$ are determined according to the measurements in recent observations of young stellar objects with infall streamers \citep{2020ApJ...904L...6A,2020NatAs...4.1158P,2024A&A...682A..61C}. For instance, \citet{2024A&A...682A..61C} observe a 4000 au streamer infalling onto the disk of a Class I protostar ([MGM2012] 512), whose infall rate is measured to be $1.5 \times 10^{-6}\,M_{\odot}\,\rm yr^{-1}$ ($\approx 0.5\,M_{\oplus}\,\rm yr^{-1}$) with a free-fall timescale of 50 kyr. In our case, assuming a streamer with a free-fall timescale of $t_{\rm ff} = \sqrt{R^3/GM_{\star}} = 50\,\text{kyr}$ and given $M_{\star} = M_{\odot}$, the length of the streamer is estimated to be $R \approx 4600\,\rm au$, which is consistent with the order of magnitude observed.

\section{Results} \label{sec:results}

We present the detailed processes of pressure bump formation, dust accumulation, and planetesimal formation using a fiducial setup. The impact of each individual parameter on gas properties and planetesimal mass evolution is analyzed using the one-factor-at-a-time method. The effects of these parameters on the success of planetesimal formation are statistically investigated through random sampling.

\subsection{Fiducial setup} \label{ssec:fidu}

\begin{figure}
\resizebox{\hsize}{!}{\includegraphics{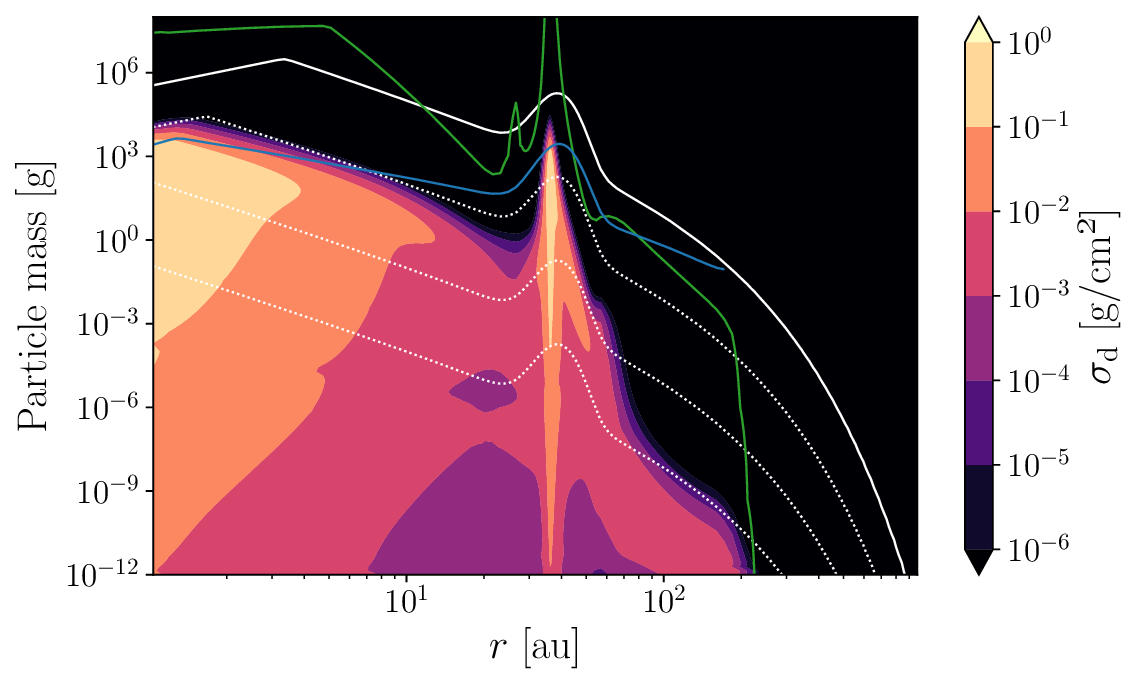}}
\caption{Dust distribution of the fiducial simulation at the start of planetesimal formation. The dust surface density $\sigma_{\rm d}$ is independent of the mass grid. The green contour indicates the drift limit, while the blue contour marks the fragmentation limit. White contours represent particle masses corresponding to Stokes numbers St = 1 (solid) and St = \{$10^{-1}$, $10^{-2}$, $10^{-3}$\} (dotted), respectively.}
\label{fig:fidu_sigma}
\end{figure}

\begin{figure}
\resizebox{\hsize}{!}{\includegraphics{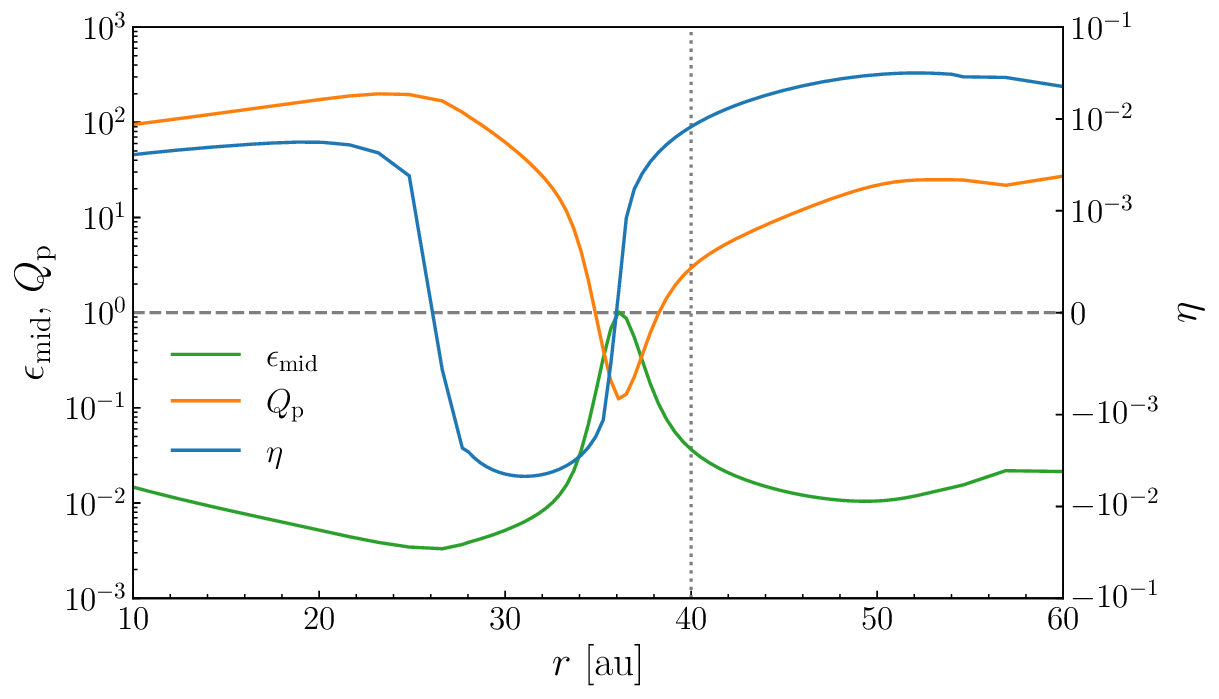}}
\caption{Midplane dust-to-gas ratio $\epsilon_{\rm mid}$, Toomre-like value $Q_{\rm p}$, and pressure gradient parameter $\eta$ around the infall region of the fiducial simulation at the start of planetesimal formation. The vertical dotted gray line indicates $r_{\rm infall} = 40$ au. The horizontal dashed gray line represents the critical conditions for streaming instability, where $\epsilon_{\rm mid} = 1$ and $Q_{\rm p} = 1$, as well as the condition for the presence of a pressure bump, $\eta = 0$.}
\label{fig:fidu_eta}
\end{figure}

The fiducial values of the parameters are displayed in Table \ref{tab:param}. In the fiducial simulation, planetesimals begin to form at $t = 1.12\times 10^5\,\rm yr$. Figure \ref{fig:fidu_sigma} shows the radial and particle mass distribution of the dust at this moment. The dust surface density $\sigma_{\rm d}$ here is defined to be independent of the particle mass grid \citep{2022ApJ...935...35S}. The drift limit, denoted by the green contour, is reached when the dust grains grow to a size where the drift timescale becomes shorter than the growth timescale \citep{2012A&A...539A.148B}, given by
\begin{equation}
    \text{St}_{\rm drift} = \frac{\epsilon}{h^2} \left|\frac{\partial \ln{P}}{\partial \ln{r}} \right|^{-1},
    \label{eq:stdr}
\end{equation}
where $\epsilon = \Sigma_{\rm d} / \Sigma_{\rm g} $ is the vertically integrated dust-to-gas ratio. The two spikes on this curve indicate the positions where $\eta$ changes sign. The fragmentation limit, denoted by the blue contour, is an approximation considering only turbulent motion, calculated as
\begin{equation}
    \text{St}_{\rm frag} = \frac{3-\sqrt{9-4b^2}}{2b},
    \label{eq:stfr}
\end{equation}
with $b = v_{\rm frag}^2/\alpha c_{\rm s}^2$. For $b \ll 1$ (i.e., $\text{St}_{\rm frag} \ll 1$), this expression simplifies to $\text{St}_{\rm frag} = b/3$. Dust growth is limited mainly by the fragmentation limit at the inner part of the disk, and by the drift limit at the outer part. The pressure gradient parameter $\eta$ shown in Figure \ref{fig:fidu_eta} has switched sign between $\sim$26 and $\sim$36 au by this time, forming a pressure bump that peaks at $\sim$36 au. The negative part of $\eta$ forces the dust in this super-Keplerian region to drift outward, traps the infalling dust, and stops the dust from the outer disk drifting inward, causing strong dust accumulation at the peak of the pressure bump. The increasing dust-to-gas ratio in the pressure bump accelerates dust growth, because the dust growth timescale given by \citep{2012A&A...539A.148B}
\begin{equation}
    t_{\rm grow} \simeq \frac{1}{\epsilon \Omega_{\rm K}}
    \label{eq:t_grow}
\end{equation}
is shortened. As shown in Figure \ref{fig:fidu_sigma}, a significant proportion of the dust in the pressure bump grows to $\text{St} > 0.1$, whose growth is constrained by the fragmentation limit. The midplane dust-to-gas ratio $\epsilon_{\rm mid}$ gradually rises and reaches the threshold 1, while the increasing $\rm St_{\rm avg}$ and $\Sigma_{\rm d,local}$ reduce $Q_{\rm p}$ below the threshold 1, as Figure \ref{fig:fidu_eta} shows. Then, streaming instability is triggered in the pressure bump, transforming the surface density of the dust into that of planetesimals.

\begin{figure}
\resizebox{\hsize}{!}{\includegraphics{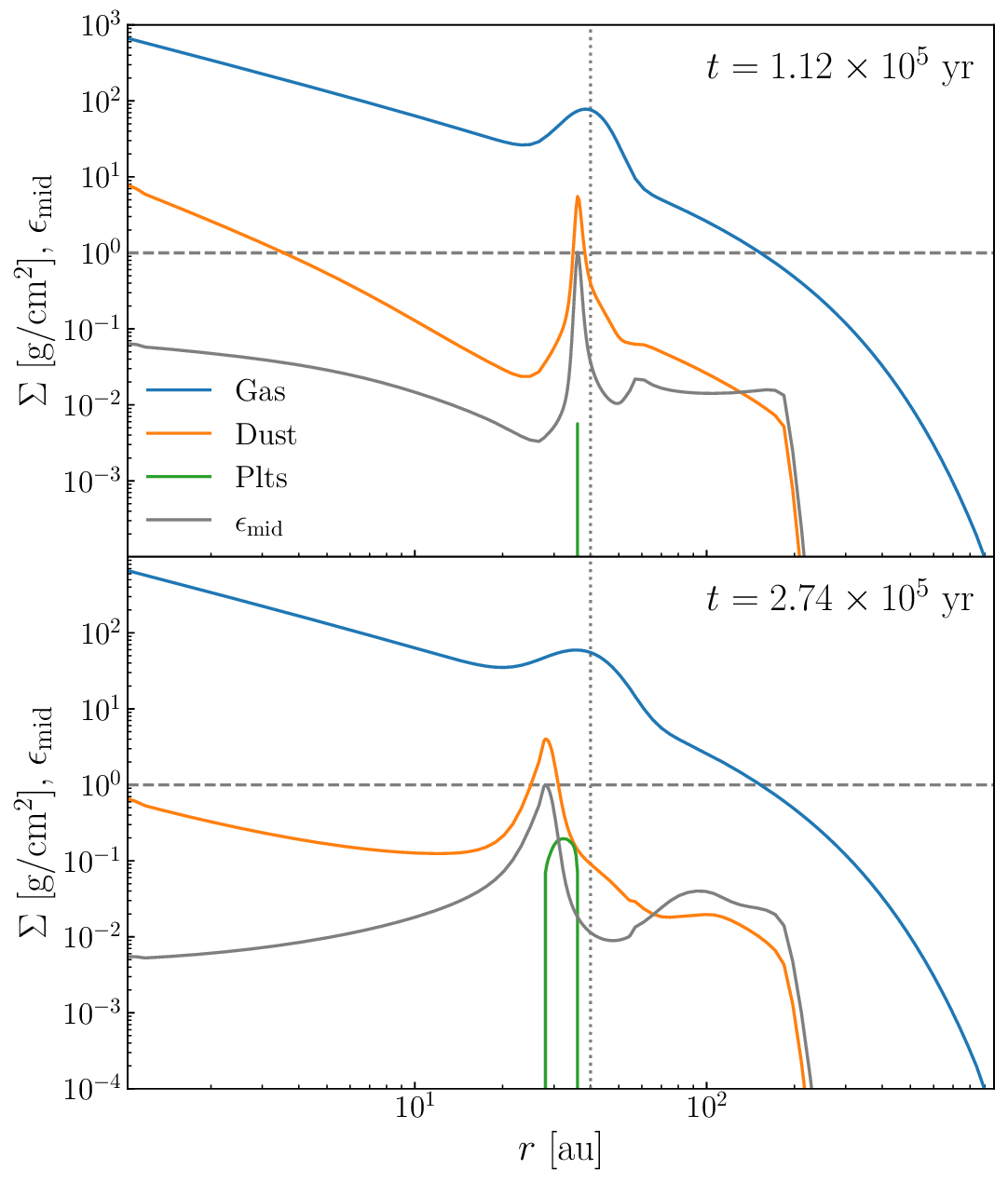}}
\caption{Radial surface density distributions of gas, dust, and planetesimals, as well as the midplane dust-to-gas ratio $\epsilon_{\rm mid}$ for the fiducial simulation at the start (top) and end (bottom) of planetesimal formation. The vertical dotted gray line indicates $r_{\rm infall} = 40$ au. The horizontal dashed gray lines represent $\epsilon_{\rm mid} = 1$.}
\label{fig:fidu_evol}
\end{figure}

Planetesimal formation stops at $t = 2.74\times 10^5\,\rm yr$. Figure \ref{fig:fidu_evol} shows the radial profiles of the gas, dust, and planetesimal surface densities as well as the midplane dust-to-gas ratio $\epsilon_{\rm mid}$, at the start and end of the planetesimal formation process, respectively. The infalling gas produces a local maximum slightly inside $r_{\rm infall} = 40\,\rm au$ in the gas surface density profile, which induces a pressure bump slightly inside the local maximum of the gas surface density $\Sigma_{\rm g}$. By the time when planetesimals start to form, the dust accumulated by the pressure bump has formed a sharp peak in the dust surface density profile, creating a similar shape for the profile of $\epsilon_{\rm mid}$. At this moment, planetesimal formation is initiated in only one radial cell where $\epsilon_{\rm mid}$ reaches 1, which is in reality a thin planetesimal ring in the protoplanetary disk. During the period of planetesimal formation, the viscous evolution of the gas gradually smooths out the bump of $\Sigma_{\rm g}$ and makes it expand, and meanwhile the peaks of the pressure bump, the dust surface density $\Sigma_{\rm d}$, and $\epsilon_{\rm mid}$ are shifted inward. At $t = 2.42\times 10^5\,\rm yr$, $\eta$ becomes positive throughout the disk, implying that the pressure trap vanishes. Then 32 kyr later, the maximum of $\epsilon_{\rm mid}$ drops below 1, thus the planetesimal formation process eventually ceases. At this point, a planetesimal belt with a width of $\sim$8 au has formed, spreading from $\sim$36 to $\sim$28 au as a result of the viscous diffusion of the pressure bump. This process closely resembles the simulation results of \citet{2021MNRAS.508.5638M}, where a migrating planet creates a moving dust trap that forms a planetesimal belt.

\begin{figure}
\resizebox{\hsize}{!}{\includegraphics{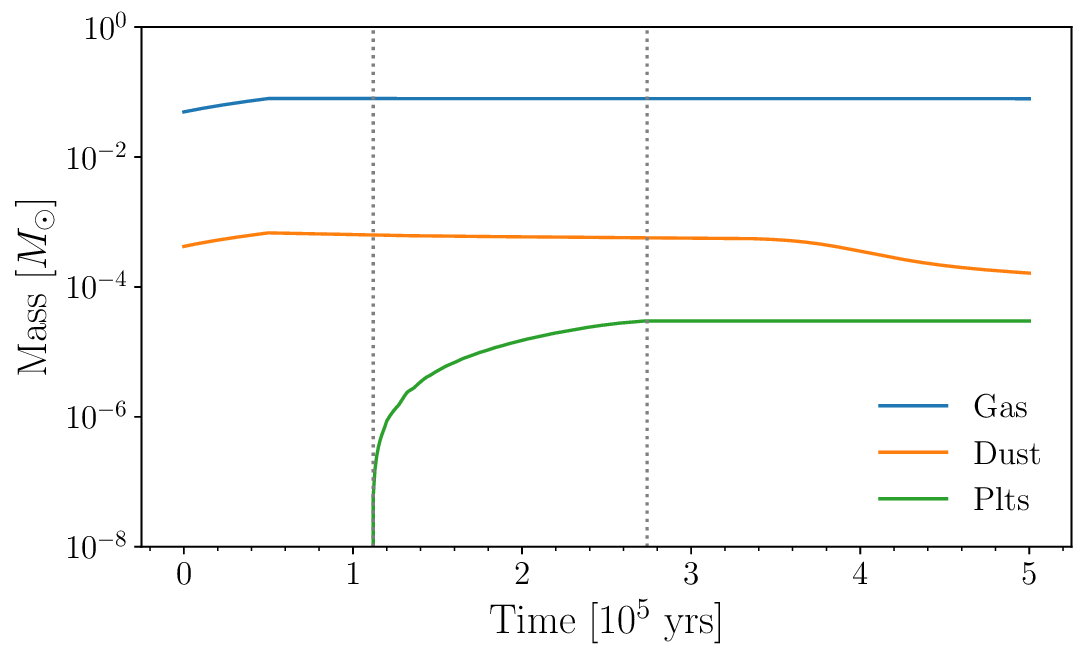}}
\caption{Mass evolution of the gas, dust, and planetesimals over 500 kyr in the fiducial simulation. Dotted lines indicate the start and end of planetesimal formation.}
\label{fig:fidu_mass}
\end{figure}

Figure \ref{fig:fidu_mass} shows the time evolution of the gas, dust, and planetesimal masses over the entire simulation. During the first 50 kyr, both the masses of the gas and dust rise due to the infall process. Over the remaining 450 kyr, the gas mass decreases slowly because of viscous accretion, while the loss of the dust mass is contributed by radial drift and transformation into planetesimals. The rapid decline of the dust mass over the last 100 kyr is because the massive dust trapped in the pressure bump that is not converted into planetesimals drifts inward after the pressure bump vanishes. Planetesimals with a total mass of 9.97 $M_{\oplus}$ formed over $1.62\times 10^5$ years, which accounts for 3.7\% of the total dust mass (the initial dust mass in the disk plus the infalling dust mass). A movie showing the result of the fiducial simulation is available online.

\subsection{One-factor-at-a-time analysis} \label{ssec:par_var}

\begin{figure*}
\centering
\includegraphics[width=17cm]{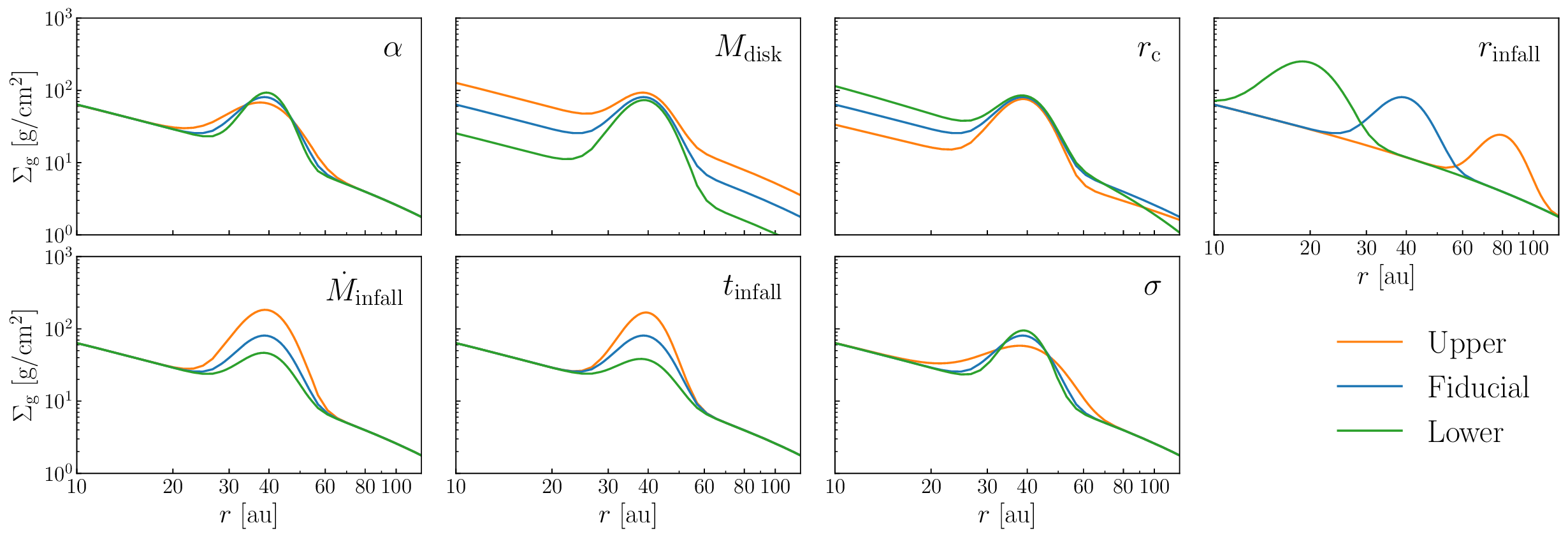}
\caption{Gas surface density distributions at $t = 100$ kyr for varying parameters compared with the fiducial setup. Each panel is labeled with the varied parameter, with green and orange curves representing decreased and increased parameter values in Table \ref{tab:param}, respectively.}
\label{fig:SigmaGas}
\end{figure*}

\begin{figure*}
\centering
\includegraphics[width=17cm]{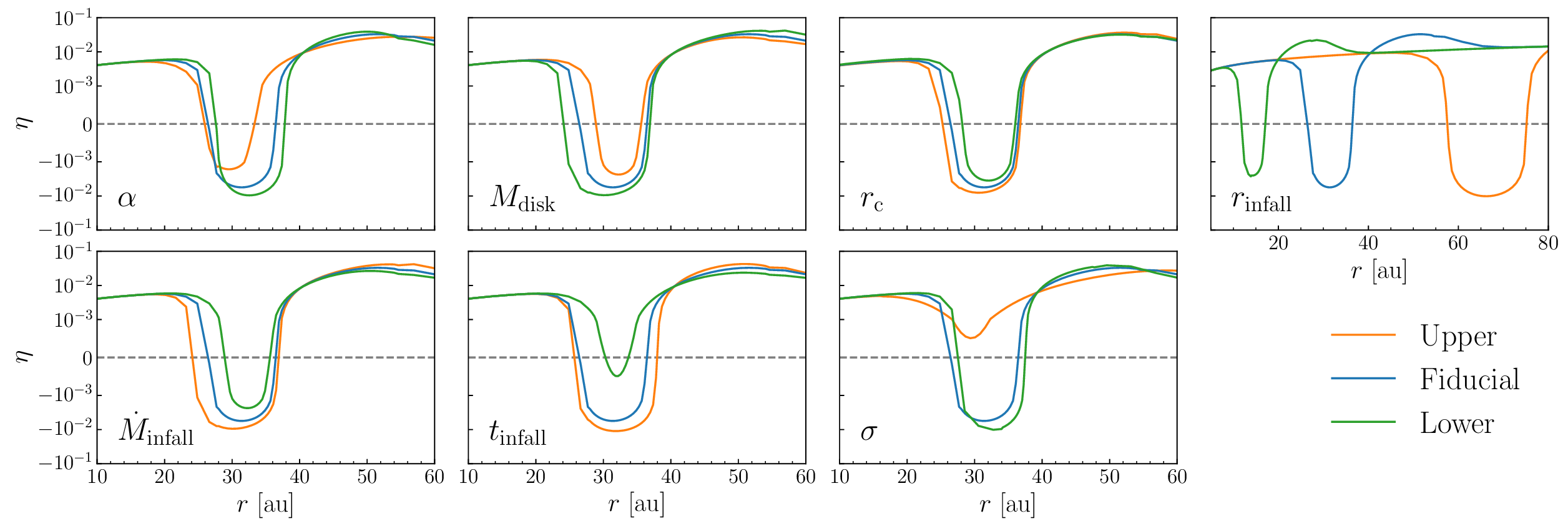}
\caption{Pressure gradient parameter $\eta$ at $t = 100$ kyr for varying parameters compared with the fiducial setup. Each panel is labeled with the varied parameter, with green and orange curves representing decreased and increased parameter values in Table \ref{tab:param}, respectively. The horizontal dashed gray lines represent $\eta = 0$.}
\label{fig:eta_all}
\end{figure*}

\begin{figure*}
\centering
\includegraphics[width=17cm]{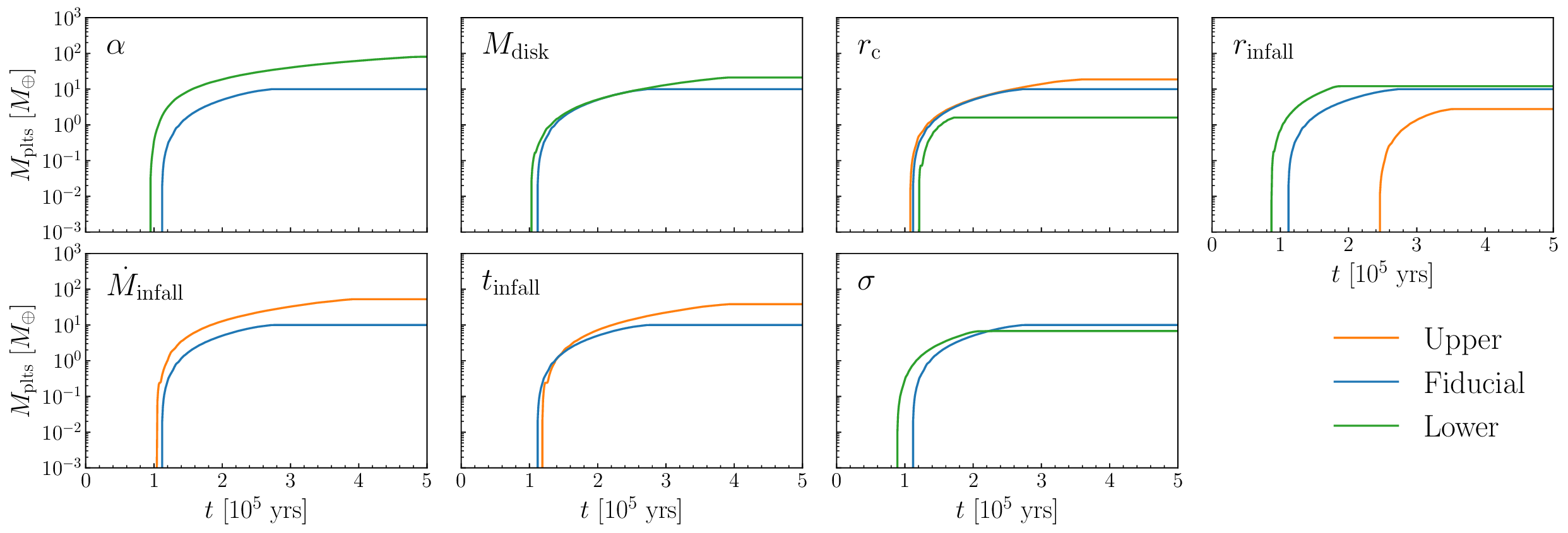}
\caption{Planetesimal mass evolution over 500 kyr for varying parameters compared with the fiducial setup. Each panel is labeled with the varied parameter, with green and orange curves representing decreased and increased parameter values in Table \ref{tab:param}, respectively. Simulations that do not produce planetesimals are not displayed.}
\label{fig:mass_all}
\end{figure*}

\begin{table*}[]
\caption{Results for one-factor-at-a-time simulations.}
\label{tab:results}
\centering
\begin{tabular}{l|c|cc|cc|cc|cc|cc|cc|cc}
\hline\hline
Parameter                     & Fid- & \multicolumn{2}{c|}{$\alpha\,(10^{-4})$} & \multicolumn{2}{c|}{$M_{\rm disk}$} & \multicolumn{2}{c|}{$r_{\rm c}$} & \multicolumn{2}{c|}{$r_{\rm infall}$} & \multicolumn{2}{c|}{$\dot{M}_{\rm infall}$} & \multicolumn{2}{c|}{$t_{\rm infall}$} & \multicolumn{2}{c}{$\sigma$}     \\ \cline{1-1} \cline{3-16} 
Value                         & ucial  & \multicolumn{1}{c|}{1}               & 4         & \multicolumn{1}{c|}{0.02}           & 0.1         & \multicolumn{1}{c|}{50}      & 200    & \multicolumn{1}{c|}{20}        & 80        & \multicolumn{1}{c|}{0.1}                    & 0.5                       & \multicolumn{1}{c|}{20}       & 100         & \multicolumn{1}{c|}{0.05}  & 0.2 \\ \hline
$t_{\rm start}$ (kyr)         & 112   & \multicolumn{1}{c|}{95}              &           & \multicolumn{1}{c|}{103}            &             & \multicolumn{1}{c|}{121}     & 108    & \multicolumn{1}{c|}{87}        & 246       & \multicolumn{1}{c|}{}                       & 104                       & \multicolumn{1}{c|}{}         & 119         & \multicolumn{1}{c|}{89}    &     \\
$t_{\rm end}$ (kyr)           & 274   & \multicolumn{1}{c|}{498}             &           & \multicolumn{1}{c|}{392}            &             & \multicolumn{1}{c|}{173}     & 356    & \multicolumn{1}{c|}{187}       & 353       & \multicolumn{1}{c|}{}                       & 391                       & \multicolumn{1}{c|}{}         & 392         & \multicolumn{1}{c|}{232}   &     \\
$M_{\rm plts}\,(M_{\oplus})$ & 9.97  & \multicolumn{1}{c|}{80.57}           & 0         & \multicolumn{1}{c|}{21.17}          & 0           & \multicolumn{1}{c|}{1.60}    & 18.67  & \multicolumn{1}{c|}{12.08}     & 2.77      & \multicolumn{1}{c|}{0}                      & 52.60                     & \multicolumn{1}{c|}{0}        & 38.20       & \multicolumn{1}{c|}{6.80}  & 0   \\
$M_{\rm plts}/M_{\rm d,tot}\,(\%)$  & 3.7 & \multicolumn{1}{c|}{30.2}          & 0         & \multicolumn{1}{c|}{12.7}         & 0           & \multicolumn{1}{c|}{0.6}   & 7.0  & \multicolumn{1}{c|}{4.5}     & 1.0     & \multicolumn{1}{c|}{0}                      & 12.6                    & \multicolumn{1}{c|}{0}        & 10.4      & \multicolumn{1}{c|}{2.6} & 0   \\
$M_{\rm plts}/M_{\rm d,out}\,(\%)$  & 5.0 & \multicolumn{1}{c|}{40.0}          & 0         & \multicolumn{1}{c|}{15.7}         & 0           & \multicolumn{1}{c|}{0.8}   & 10.1  & \multicolumn{1}{c|}{5.5}     & 1.6     & \multicolumn{1}{c|}{0}                      & 15.0                    & \multicolumn{1}{c|}{0}        & 12.7      & \multicolumn{1}{c|}{3.5} & 0   \\ \hline
\end{tabular}
\tablefoot{The start and end time of planetesimal formation, the total mass of planetesimals formed, and the conversion efficiency from dust to planetesimals are presented. The parameters use the same units as those listed in Table \ref{tab:param}. $M_{\rm d,tot}$ represents the total dust mass, comprising both the dust initially in the disk and the dust in the infalling material. $M_{\rm d,out}$ denotes the dust mass in the outer disk available for trapping, including both the infalling dust and the dust initially in the disk outside $(r_{\rm infall} - 3 \sigma_{r})$.}
\end{table*}

In each simulation, only one parameter --- except for $v_{\rm frag}$ (as it does not affect gas properties) --- is varied to its maximum or minimum value from the ranges listed in Table \ref{tab:param}, while all other parameters are held at their fiducial values. The resulting distributions of gas surface density and the midplane pressure gradient parameter $\eta$ at 100 kyr for these simulations are illustrated in Figures \ref{fig:SigmaGas} and \ref{fig:eta_all}, respectively. Figure \ref{fig:mass_all} presents the evolution of planetesimal mass in the simulations where planetesimal formation occurs. We select $t = 100$ kyr as the snapshot time because, by this point, the infall stages have concluded across all simulations. Furthermore, in most simulations where planetesimals form (except when $r_{\rm infall} = 80,\rm au$), the trigger time for planetesimal formation occurs around $t = 100$ kyr, as shown in Figure \ref{fig:mass_all}. In simulations that do not produce planetesimals, this time is close to the point when their $\epsilon_{\rm mid}$ values reach their maximum, though still below 1. Table \ref{tab:results} summarizes the simulation outcomes, including the start and end times of planetesimal formation, the total planetesimal masses formed, and the conversion efficiencies from dust to planetesimals.

For the viscosity parameter $\alpha$, the results for $\alpha = 1 \times 10^{-4}$ (green) and $\alpha = 4 \times 10^{-4}$ (orange) are presented. With the same amount of gas dumped onto identical disks, the Gaussian-like gas surface density profile is broader and lower for higher $\alpha$ compared to lower $\alpha$. The pressure gradient parameter $\eta$ is correlated with the slope of the gas surface density profile by the relations in Eqs. (\ref{eq:rho_g}) and (\ref{eq:eta}). Consequently, the region with negative $\eta$ spans a narrower range of $r$ for higher $\alpha$, and has lower absolute values of $\eta$. For $\alpha = 1\times 10^{-4}$, the planetesimal formation process is much longer-lasting and more productive than that of the fiducial setup. In contrast, no planetesimal is formed in the simulation with $\alpha = 4\times 10^{-4}$ since $\epsilon_{\rm mid}$ never reaches 1, though the condition for $Q_{\rm p}$ can be satisfied.

For the initial disk mass $M_{\rm disk}$, the results for $M_{\rm disk} = 0.02\,M_{\odot}$ (green) and $M_{\rm disk} = 0.1\,M_{\odot}$ (orange) are presented. Although the amount of infalling gas is the same, the Gaussian-like gas surface density profiles differ in height due to the varying initial profiles of the disks. For smaller $M_{\rm disk}$, the bump in the gas surface density profile is more prominent, leading to a steeper slope and consequently a higher and broader absolute negative pressure gradient. For $M_{\rm disk} = 0.02\,M_{\odot}$, the planetesimal formation process is longer-lasting and more productive compared to the fiducial setup. In contrast, no planetesimal is formed for $M_{\rm disk} = 0.1\,M_{\odot}$ since $\epsilon_{\rm mid}$ never reaches 1, though the condition for $Q_{\rm p}$ can be satisfied.

For the characteristic radius of disk $r_{\rm c}$, the results for $r_{\rm c} = 50\,\rm au$ (green) and $r_{\rm c} = 200\,\rm au$ (orange) are presented. Since $r_{\rm infall} = 40 \,\rm au$ is always smaller than $r_{\rm c}$ here, the initial gas surface density is higher in the infall region for smaller $r_{\rm c}$, resulting in a flatter gas surface density bump. Hence, the region with negative $\eta$ is broader and has higher absolute values of $\eta$ for larger $r_{\rm c}$. The planetesimal mass produced with $r_{\rm c} = 50\,\rm au$ is much less than that of the fiducial setup, while the total planetesimal mass of $r_{\rm c} = 200\,\rm au$ is approximately twice that of the fiducial setup.

For the central infall radius $r_{\rm infall}$, the results for $r_{\rm infall} = 20\,\rm au$ (green) and $r_{\rm infall} = 80\,\rm au$ (orange) are presented. We note that the horizontal scale of the panel of $r_{\rm infall}$ in Figure \ref{fig:eta_all} is different from the other panels in the same Figure. Due to constant $\sigma$, the widths of the gas surface density bumps (as well as the pressure bumps) increase with $r_{\rm infall}$. The absolute values of negative $\eta$ also increase with $r_{\rm infall}$, because the initial gas surface density profile drops with increasing $r$. For $r_{\rm infall} = 20\,\rm au$, planetesimal formation starts significantly earlier than the fiducial setup and produces a bit more planetesimal mass than the fiducial setup. For $r_{\rm infall} = 80\,\rm au$, planetesimal formation starts much later than the fiducial setup, and produces much less planetesimal mass than the fiducial setup.

For the gas infall rate $\dot{M}_{\rm infall}$, the results for $\dot{M}_{\rm infall} = 0.1\, M_{\oplus}\,\rm yr^{-1}$ (green) and $\dot{M}_{\rm infall} = 0.5\, M_{\oplus}\,\rm yr^{-1}$ (orange) are presented. With the same initial disk conditions, the Gaussian-like gas surface density profile is higher and has a higher slope for larger $\dot{M}_{\rm infall}$ than smaller $\dot{M}_{\rm infall}$, resulting in a broader region of negative $\eta$ with higher absolute values. For $\dot{M}_{\rm infall} = 0.1\, M_{\oplus}\,\rm yr^{-1}$, no planetesimal is formed since $\epsilon_{\rm mid}$ never reaches 1, though the condition for $Q_{\rm p}$ can be satisfied. For $\dot{M}_{\rm infall} = 0.5\, M_{\oplus}\,\rm yr^{-1}$, the results are quite similar to those of the simulation with $M_{\rm disk} = 0.02\,M_{\odot}$ except the total planetesimal mass. What the two systems have in common besides other parameters is the mass ratio between the infalling gas and the gaseous disk $(\dot{M}_{\rm infall} t_{\rm infall})/M_{\rm disk}$. The difference in total planetesimal masses of the two systems is caused by different dust mass budgets.

For the infall duration $t_{\rm infall}$, the results for $t_{\rm infall} = 20\,\rm kyr$ (green) and $t_{\rm infall} = 100\,\rm kyr$ (orange) are presented. Since the infall rate $\dot{M}_{\rm infall}$ is constant, $t_{\rm infall}$ determines the total mass of infalling gas, leading to a higher gas surface density profile for larger $t_{\rm infall}$. For $t_{\rm infall} = 20\,\rm kyr$, the lengthy evolution time after the end of infall until $t = 100$ kyr flattens the gas surface density profile and makes the region with negative $\eta$ negligible at this point. No planetesimal is formed during the evolution process since $\epsilon_{\rm mid}$ never reaches 1, though the condition for $Q_{\rm p}$ can be satisfied. For $t_{\rm infall} = 100\,\rm kyr$, the planetesimal formation process is much longer-lasting and more productive compared to the fiducial setup, though it starts slightly later.

For the relative infall width $\sigma$, the results for $\sigma = 0.05$ (green) and $\sigma = 0.2$ (orange) are presented. The gas surface density profiles present a similar trend as varying $\alpha$: it is broader and lower for higher $\sigma$. For $\sigma = 0.05$, the negative part of $\eta$ has higher absolute values than the fiducial setup, but spans a narrower range of $r$. The planetesimal formation process is longer-lasting than that of the fiducial setup, but less productive. For $\sigma = 0.2$, $\eta$ is positive everywhere at this point, meaning that there is no pressure bump. Actually, a negligible pressure bump formed at 41 kyr and vanished at 74 kyr, but eventually no planetesimal is formed throughout the simulation since $\epsilon_{\rm mid}$ never reaches 1. However, the condition for $Q_{\rm p}$ was still achieved in the simulation.

\subsection{Sobol sampling analysis} \label{ssec:sobol}

\begin{figure*}[htbp]
\sidecaption
\includegraphics[width=12cm]{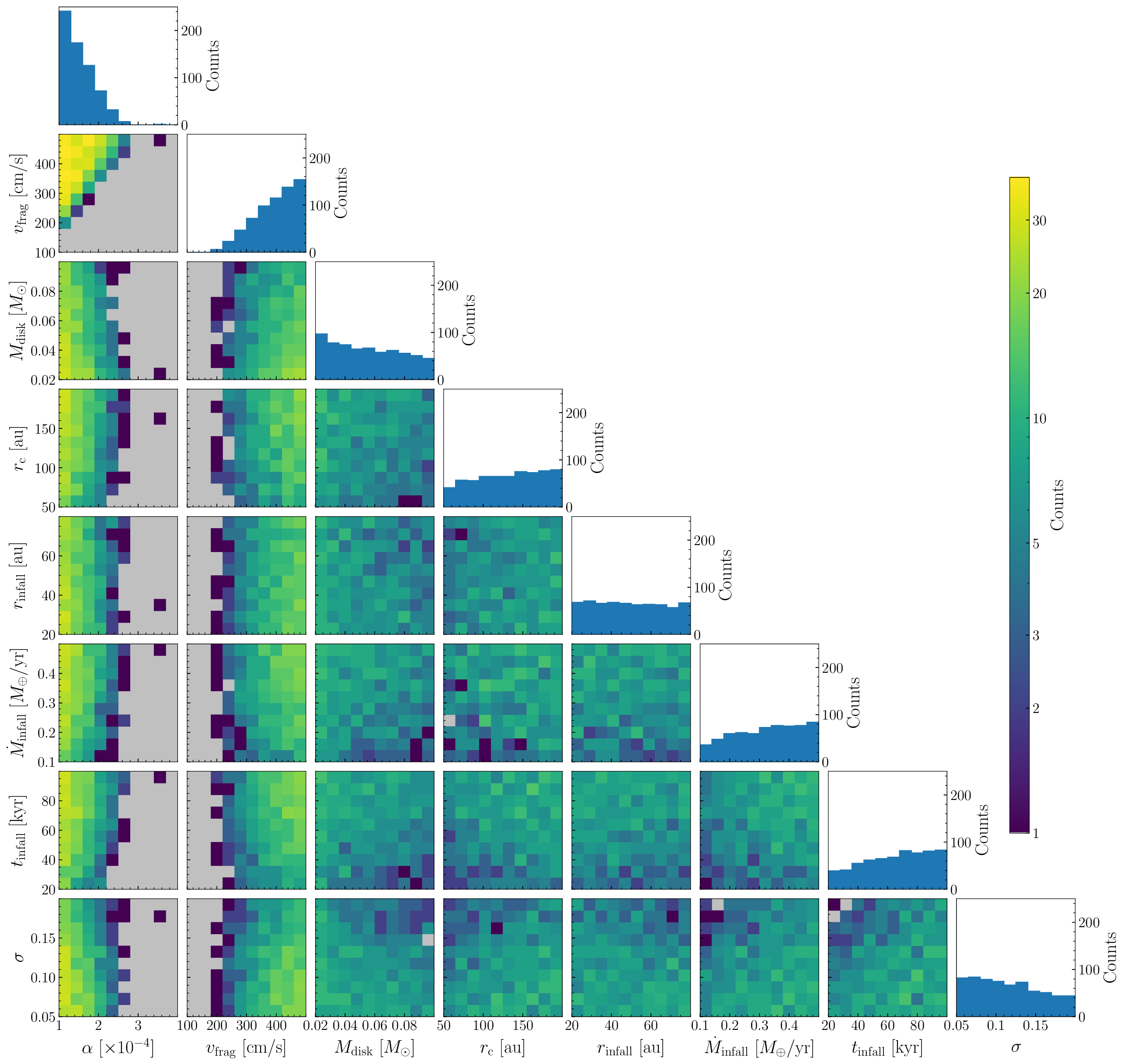}
\caption{Histograms of 653 simulations where planetesimal formation occurs, out of 4096 Sobol-sampled simulations. "Counts" refers to the number of simulations. The 1D histograms along the diagonal show the distribution over each of the eight parameters individually. The 2D heatmap histograms display the joint distributions over pairs of parameters, with gray cells indicating a count of 0.}
\label{fig:sobol_results}
\end{figure*}

We generated 4096 simulation setups, uniformly sampling the eight parameters within the value ranges shown in Table \ref{tab:param}. A Sobol sequence \citep{SOBOL196786} was used to achieve a quasi-random yet unclustered selection of parameter sets. Of the 4096 simulations, 653 successfully formed planetesimals. The histograms in Figure \ref{fig:sobol_results} illustrate the statistical distributions of these successful simulations across the eight parameters, and each parameter is divided into ten equal bins within its value range.

Among all parameters, the simulation outcomes are most sensitive to $\alpha$, as evidenced by the steepest gradients in the first column of Figure \ref{fig:sobol_results} compared to the other parameters. At $\alpha \approx 10^{-4}$, about 60\% of the simulations result in planetesimal formation. This fraction drops rapidly as $\alpha$ increases. Simulations with $\alpha > 2.5\times10^{-4}$ rarely produce planetesimals. The viscous diffusion timescale of the gas across the infall-induced pressure bump can be expressed as
\begin{equation}
    t_{\rm visc} \simeq \frac{\sigma_{r}^2}{\nu} = \frac{\sigma^2 r_{\rm infall}^2}{\alpha c_{\rm s} H_{\rm g}}.
    \label{eq:t_visc}
\end{equation}
A higher $\alpha$ leads to increased viscosity and a shorter gas diffusion timescale, which causes the pressure bump to smooth out more quickly. Conversely, with lower $\alpha$, the pressure bump is retained longer, providing more time for dust to accumulate, grow, and trigger streaming instability.

The second most decisive parameter is $v_{\rm frag}$. Simulations with $v_{\rm frag} < 200\,\rm cm\,s^{-1}$ rarely form planetesimals. As $v_{\rm frag}$ increases, the maximum Stokes number of dust particles, determined by Eq. (\ref{eq:stfr}), also rises, leading to a higher average St for all particles. According to Eq. (\ref{eq:H_d}), particles with larger St are vertically distributed across a lower dust scale height $H_{\rm d}$. Consequently, a higher average St increases the midplane dust density, $\rho_{\rm d}$. This raises the midplane dust-to-gas ratio, $\epsilon_{\rm mid} = \rho_{\rm d}/\rho_{\rm g}$, facilitating the attainment of the critical condition for streaming instability, $\epsilon_{\rm mid} = 1$.

For $M_{\rm disk}$ and $r_{\rm c}$, two parameters describing disk characteristics, the key to their effects on planetesimal formation is the initial gas mass within the infall region. According to Eq. (\ref{eq:init}), $M_{\rm disk}$ uniformly scales the magnitude of the initial gas surface density profile, while $r_{\rm c}$ determines how the initial gas mass is distributed across the disk radius $r$. A smaller $M_{\rm disk}$ reduces $\Sigma_{\rm g,init}$, which leads to a steeper pressure bump under identical infall conditions. As $r_{\rm c}$ increases while $M_{\rm disk}$ remains fixed, the gas surface density decreases in the inner region of the disk and increases in the outer region, resulting in a flatter and more spatially extended profile. Since the range of $r_{\rm infall}$ tested is of the same order of magnitude as the minimum $r_{\rm c}$, the infall region is located within the inner part of the disk. A larger $r_{\rm c}$ is conducive to forming the pressure bump, as it results in a lower initial gas surface density $\Sigma_{\rm g,init}$ at $r_{\rm infall}$. If $r_{\rm infall}$ is located at hundreds of au, the effect of $r_{\rm c}$ will become more complex.

The infall location, $r_{\rm infall}$, exhibits the weakest impact on the simulation outcomes among all parameters. According to Eq. (\ref{eq:t_grow}), a lower orbital frequency $\Omega_{\rm K}$ at a larger $r_{\rm infall}$ leads to a longer dust growth timescale, which hinders and postpones the onset of planetesimal formation. Nevertheless, the lower $\Sigma_{\rm g,init}$ at a larger $r_{\rm infall}$ promotes the formation of the pressure bump, similar to the effects of a smaller $M_{\rm disk}$ and a larger $r_{\rm c}$. Consequently, the favorable and unfavorable factors for planetesimal formation at larger $r_{\rm infall}$ tend to offset each other. Although the histograms in Figure \ref{fig:sobol_results} show a slight preference for smaller $r_{\rm infall}$, this parameter has a negligible impact on the likelihood of planetesimal formation and primarily influences its timing.

The other parameters describing infall properties, $\dot{M}_{\rm infall}$, $t_{\rm infall}$, and $\sigma$, influence planetesimal formation by modulating the amount of infalling material and its distribution. The effects of $\dot{M}_{\rm infall}$ and $t_{\rm infall}$ are primarily determined by their product, $M_{\rm infall}$, which represents the total mass of the infalling gas. This is evidenced by the 2D histogram of $\dot{M}_{\rm infall}$ and $t_{\rm infall}$ in Figure \ref{fig:sobol_results}. Since only two of these three quantities are independent, we selected the two that are mostly measured in recent observational studies of infall streamers \citep{2020ApJ...904L...6A,2020NatAs...4.1158P,2024A&A...682A..61C}. A larger $\dot{M}_{\rm infall}$ or $t_{\rm infall}$ leads to a greater $M_{\rm infall}$, which not only enhances the prominence of the pressure bump, but also provides more dust material to build planetesimals. The relative width of the infall region, $\sigma$, determines how concentrated $M_{\rm infall}$ is distributed over $r$. A lower $\sigma$ creates a steeper but narrower pressure bump. While the steepness aids in triggering streaming instability, the narrower width limits the amount of material available for planetesimal formation, which explains the intersection of the two curves in the panel of $\sigma$ in Figure \ref{fig:mass_all}.

In summary, planetesimal formation in an infall-induced pressure bump, with $r_{\rm infall}$ located within a few tens of au, is more likely to occur with lower values of $\alpha$, $M_{\rm disk}$, and $\sigma$, and higher values of $v_{\rm frag}$, $r_{\rm c}$, $\dot{M}_{\rm infall}$, and $t_{\rm infall}$. The most decisive two factors are $\alpha$ and $v_{\rm frag}$, as planetesimal formation becomes highly improbable when $\alpha > 2.5 \times 10^{-4}$ and $v_{\rm frag} < 200\,\rm cm\,s^{-1}$. The ideal scenario is a low-viscosity, low-mass, but spatially extended disk accreting a massive yet narrow infall streamer.

\section{Discussions} \label{sec:dis}

\subsection{Timescale ratios} \label{ssec:ratio}

\begin{figure}
\resizebox{\hsize}{!}{\includegraphics{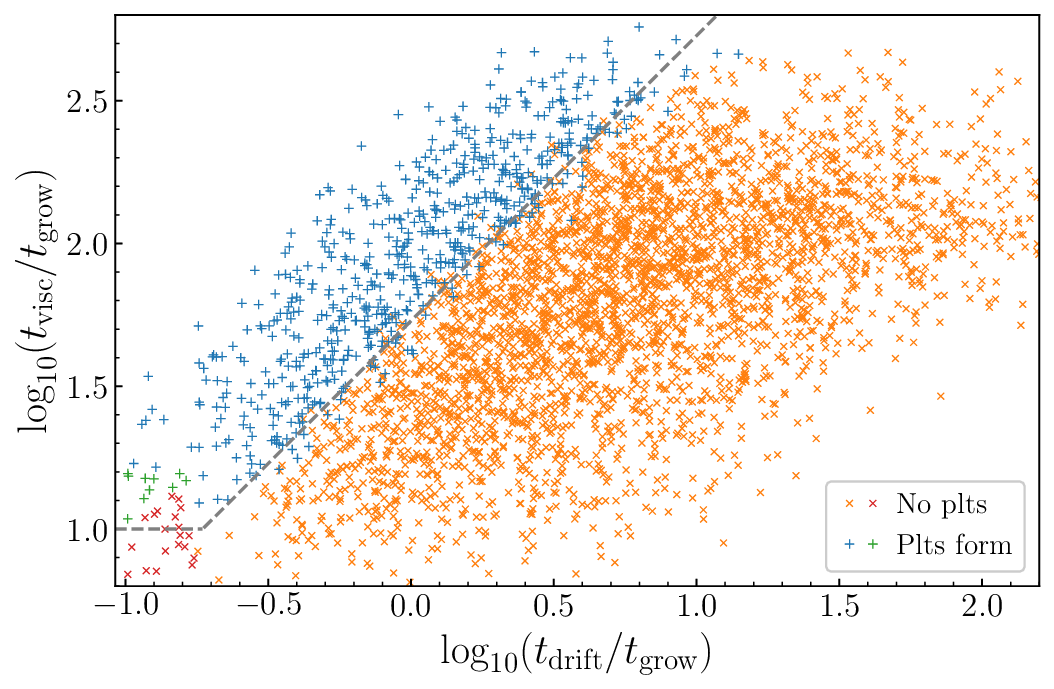}}
\caption{Distribution of timescale ratios $t_{\rm drift}/t_{\rm grow}$ and $t_{\rm visc}/t_{\rm grow}$. The plus symbols ("+") represent simulations in which planetesimals form, while the cross symbols ("x") indicate simulations with no planetesimal formation. Blue and orange markers show the Sobol-sampled simulations, while red and green markers represent a subset of 30 selected simulations from these Sobol-sampled simulations, with a reduced initial global dust-to-gas ratio ($\epsilon_{\rm init} = 10^{-2.25}$). The dashed gray line denotes the fitted boundary of the planetesimal-forming domain.}
\label{fig:t_ratios}
\end{figure}

In order to reach the critical conditions for planetesimal formation, a protoplanetary disk undergoing infall must satisfy the following conditions: (1) forming a pressure bump prominent enough to trap and accumulate dust; and (2) ensuring that the pressure bump persists long enough to allow dust to grow until the onset of streaming instability. Each of these two conditions can be characterized by a timescale ratio, which is calculated by the initial parameter setups of the simulations.

The presence of a pressure bump is signified by a super-Keplerian region with positive pressure gradient and negative $\eta$ values, located on the inner side of the bump. According to Eq. (\ref{eq:drift}), the minimum value of $\eta$ (i.e., the largest absolute value for negative $\eta$) corresponds to the fastest outward drift speed of dust, reflecting the pressure bump's ability to accumulate and trap dust. The dust drift timescale at this site is formulated as
\begin{equation}
    t_{\rm drift} = \frac{r}{v_{\rm d,max}} = -\frac{1 + \mathrm{St}^2}{2 \eta_{\rm min} \mathrm{St}\, \Omega_{\rm K}}.
\end{equation}
Given the disk parameters $M_{\rm disk}$ and $r_{\rm c}$, along with the infall parameters $\dot{M}_{\rm infall}$, $t_{\rm infall}$, $r_{\rm infall}$, and $\sigma$, the gas surface density profile after the infall process, assuming no viscosity, is calculated as
\begin{equation}
    \Sigma_{\rm g} = \Sigma_{\rm g,init} + S_{\rm g,infall} t_{\rm infall}.
    \label{eq:no_diff}
\end{equation}
Using the relations from Eqs. (\ref{eq:h}) to (\ref{eq:eta}), the analytical radial profile of $\eta$ can be derived, allowing for the calculation of its minimum value $\eta_{\rm min}$. The Stokes number at the fragmentation limit at this location, $\rm St_{frag}$, given by Eq. (\ref{eq:stfr}), is used in the calculation. If no fragmentation limit is present, $\rm St_{frag} = 1$ is assumed, representing the largest particles participating in streaming instability, as described in Section \ref{ssec:plts}. Given the dust growth timescale in Eq. (\ref{eq:t_grow}), the timescale ratio of drift to growth is expressed as
\begin{equation}
    \frac{t_{\rm drift}}{t_{\rm grow}} = -\frac{(1 + \mathrm{St}_{\rm frag}^2) \epsilon}{2 \eta_{\rm min} \mathrm{St}_{\rm frag}},
\end{equation}
where a constant initial global dust-to-gas ratio $\epsilon = \epsilon_{\rm init} =  0.01$ is assumed, since $\epsilon$ evolves over time and varies with $r$. This quantity is influenced by all the eight major parameters.

The lifetime of a pressure bump is measured by the gas viscous diffusion timescale across the bump, given by Eq. (\ref{eq:t_visc}). The timescale ratio of gas viscous diffusion to dust growth is formulated as
\begin{equation}
    \frac{t_{\rm visc}}{t_{\rm grow}} = \frac{\sigma^2 r_{\rm infall}^2 \Omega_{\rm K}^2 \epsilon}{\alpha c_{\rm s}^2} \propto \frac{\sigma^2}{\alpha r_{\rm infall}^{1/2}},
\end{equation}
where it is considered that $c_{\rm s}$ and $\Omega_{\rm K}$ vary only with $r_{\rm infall}$. Comparing $t_{\rm visc}$ to $t_{\rm grow}$ reveals the extent to which dust particles can grow before the pressure bump dissipates. In contrast, comparing $t_{\rm drift}$ to $t_{\rm grow}$ serves more for mathematical clarity than physical necessity, as $t_{\rm drift}$ alone effectively measures the pressure bump's ability to trap dust.

Figure \ref{fig:t_ratios} displays the distribution of the two timescale ratios in logarithmic scale, calculated for our simulations. The blue and orange markers represent the Sobol-sampled simulations, excluding cases where the pressure bump fails to form (where $t_{\rm drift}$ is negative). The two timescale ratios are not independent: a lower $\alpha$ increases $\mathrm{St}_{\rm frag}$, leading to a lower $t_{\rm drift}/t_{\rm grow}$ and a higher $t_{\rm visc}/t_{\rm grow}$; a lower $\sigma$ steepens the pressure gradient (increasing $|\eta_{\rm min}|$), which reduces both $t_{\rm drift}/t_{\rm grow}$ and $t_{\rm visc}/t_{\rm grow}$. A distinct boundary separates simulations that form planetesimals (blue) from those that do not (orange). For a given $t_{\rm visc}/t_{\rm grow}$, simulations with faster outward dust drift in the pressure bump (lower $t_{\rm drift}/t_{\rm grow}$) are more likely to produce planetesimals. Likewise, for a given $t_{\rm drift}/t_{\rm grow}$, planetesimal formation is favored by longer diffusion timescales, indicating a longer-lived pressure bump relative to dust growth timescales. A logistic regression fit gives the boundary as:
\begin{align}
    \log_{10} \left(\frac{t_{\rm visc}}{t_{\rm grow}}\right) &> 1.00 \log_{10} \left(\frac{t_{\rm drift}}{t_{\rm grow}}\right) + 1.73, \label{eq:boundary}\\
    t_{\rm visc} &\gtrsim 53.55\, t_{\rm drift},
    \label{eq:visc_grow}
\end{align}
which defines the conditions for planetesimal formation, highlighting the need for a long-lasting pressure bump that is sufficiently prominent to facilitate rapid dust accumulation.

The distribution of blue and orange markers leaves an empty region in the lower-left corner of Figure \ref{fig:t_ratios}. To investigate whether there is a lower limit for $t_{\rm visc}/t_{\rm grow}$ within the planetesimal-forming domain, we selected simulations in the range $-0.75 < \log_{10}(t_{\rm drift}/t_{\rm grow}) < -0.5$ and $1.05 < \log_{10}(t_{\rm visc}/t_{\rm grow}) < 1.45$, where most simulations form planetesimals. We then reduced their initial global dust-to-gas ratio to $\epsilon_{\rm init} = 10^{-2.25}$, filling the previously blank area. These additional simulations are shown as green and red markers in Figure \ref{fig:t_ratios}. Although many of the red crosses lie to the upper left of the boundary line given by Eq. (\ref{eq:boundary}), they do not result in planetesimal formation. This suggests that if the pressure bump's lifetime is insufficient relative to the dust growth timescale, planetesimals cannot form in time, even if dust accumulation via outward drift in the pressure bump is sufficiently rapid. We approximate this boundary as:
\begin{equation}
    t_{\rm visc} \gtrsim 10\, t_{\rm grow}.
\end{equation}
Combining this with Eq. (\ref{eq:visc_grow}) provides the critical conditions for planetesimal formation in an infall-induced pressure bump.

\subsection{Comparison of contributions from primordial dust and infalling dust}

\begin{table}[]
\caption{Total mass of planetesimals (in Earth masses) formed in simulations with only infalling dust, only dust initially in the disk, and both dust sources, at various $\alpha$ values.}
\label{tab:separate}
\centering
\begin{tabular}{l|ccc}
\hline\hline
$\alpha\,(\times 10^{-4})$ & Infall-Only & Disk-Only & Infall + Disk \\
\hline
1                   & 43.72       & 13.18     & 80.57         \\
1.5                 & 13.64       & 0         & 32.74         \\
2                   & 0           & 0         & 9.97         \\ \hline
\end{tabular}
\end{table}

In our models, the infalling streamer shares the same global dust-to-gas ratio as the initial protoplanetary disk, meaning that the solid material forming planetesimals originates from both the dust in the original disk and the infalling dust. To determine whether planetesimals can form with only one dust source, and which source contributes more, we simulated two extreme scenarios: one with a dust-free initial disk and another with a dust-free infall streamer. This experiment was conducted with $\alpha = \{1, 1.5, 2\}\times10^{-4}$, while keeping all other parameters at their fiducial values.

The results are summarized in Table \ref{tab:separate}. For $\alpha = 10^{-4}$, planetesimal formation occurs in both scenarios. The combined planetesimal mass in these two extreme scenarios is less than that of the simulation with both dust sources present. For $\alpha = 1.5 \times 10^{-4}$, planetesimal formation occurs only in the simulation with an initially dust-free disk, where the planetesimal mass formed from infalling dust alone is less than in the simulation with both dust sources. For $\alpha = 2 \times 10^{-4}$, planetesimal formation does not occur in either extreme scenario. Therefore, planetesimal formation with only one dust source is possible when conditions such as a low $\alpha$ value are particularly favorable, with the infalling dust contributing more significantly than the primordial dust in the disk.

Once an infall-induced pressure bump forms, all incoming infalling dust is trapped by the bump, whereas only a portion of the dust initially in the disk can be captured. Since the mass of the infalling dust is comparable to the total initial dust mass in the disk, the infalling dust becomes the primary contributor to planetesimal formation. However, the total amount of planetesimals formed depends on multiple factors beyond just the amount of available dust. The standard scenario, which includes both dust sources, not only traps more dust in the pressure bump but also reaches the critical conditions for streaming instability earlier than the two extreme scenarios with only one dust source. This explains why the standard scenario produces more planetesimal mass than the combined total of the two extreme scenarios.

In all the simulations above, infall begins at $t = 0$. We also tested scenarios where infall starts later. Since dust growth begins from sub-micron-sized particles, a delayed infall means that dust grains initially in the disk have already grown to larger sizes when the pressure bump forms, resulting in faster dust drift and trapping in the pressure bump. However, if infall starts too late, a significant portion of the primordial dust drifts inward to the inner disk in the absence of a dust trap, reducing the amount of dust available to be trapped by the pressure bump. We tested starting infall at 50 kyr and 100 kyr on the fiducial setup. In the former case, the total planetesimal mass increases to 11.57 $M_{\oplus}$ due to the faster accumulation of the primordial dust. In the latter, it drops to 9.71 $M_{\oplus}$ due to the reduced amount of trapped primordial dust. This effect is minor since the infalling dust plays a dominant role, and the timing of infall only affects the distribution of the dust initially present in the disk.

\subsection{Effects of other parameters}

Apart from the eight major parameters we previously discussed, two other parameters also play roles in shaping the planetesimal formation process: the small-scale diffusion parameter $\delta$ (Eq. \ref{eq:Qp}) and the planetesimal formation efficiency $\zeta$ (Eq. \ref{eq:plts_dt}). These parameters were excluded from our parameter space because they are not directly relevant to the large-scale evolution of gas and dust in the disk.

The small-scale diffusion parameter $\delta$ is defined as the ratio of radial particle diffusivity to $c_{\rm s} H_{\rm g}$ \citep{2018ApJ...861...47S}, analogous to the definition of $\alpha$ in terms of gas viscosity. According to Eq. (\ref{eq:Qp}), a decrease in $\delta$ leads to a lower Toomre-like parameter $Q_{\rm p}$, which facilitates reaching the critical condition for streaming instability, $Q_{\rm p} < 1$. For all simulations presented above, we used $\delta = 10^{-4}$, the most conservative value for planetesimal formation according to the measurements by \citet{2018ApJ...861...47S}. In contrast, recent work by \citet{2024A&A...688A..22L}, which employs the same criterion for planetesimal formation, adopts $\delta = 10^{-5}$. However, our simulation results indicate that the other criterion, $\epsilon_{\rm mid} \geq 1$, is always more restrictive than $Q_{\rm p} < 1$. Thus, reducing $\delta$ does not affect the triggering of streaming instability; its only effect in this context is to increase the activation function $\mathcal{P}_{\rm pf}$ through the reduced $Q_{\rm p}$, as shown in Eq. (\ref{eq:act}), ultimately influencing the total planetesimal mass formed, as described by Eq. (\ref{eq:plts_dt}). We tested $\delta = 10^{-5}$ on the fiducial setup and found that the start and end times of planetesimal formation were identical to those of the fiducial setup, with a slight increase in total planetesimal mass, from 9.97 $M_{\oplus}$ to 9.99 $M_{\oplus}$. Besides determining $Q_{\rm p}$, $\delta$ also plays a key role in defining the initial mass function of planetesimals, as given by \citet{2023ApJ...949...81G}. This aspect, however, is not addressed in this work, as we do not convert planetesimal surface density into planetary bodies.

The planetesimal formation efficiency $\zeta$ also has no influence on the triggering of streaming instability. As shown in Eq. (\ref{eq:plts_dt}), $\zeta$ determines the mass of planetesimals formed per unit time, with higher values of $\zeta$ resulting in faster planetesimal formation. \citet{2016A&A...594A.105D}, who propose this prescription for planetesimal transformation, demonstrate that a range of $\zeta$ values between $10^{-6}$ and $10^{-2}$ is reasonable. In all simulations presented so far, we adopted $\zeta=10^{-3}$. We tested $\zeta=10^{-2}$ on the fiducial setup. The total planetesimal mass formed is 24.67 $M_{\oplus}$, significantly higher than in the fiducial simulation. While the onset time of planetesimal formation remains the same as in the fiducial setup, the end time occurs earlier, shortening the planetesimal formation duration from 162 kyr to 119 kyr. Since planetesimal formation ceases when $\epsilon_{\rm mid}$ drops below one, a faster rate of dust consumption shortens the formation duration. Within a given time frame, a quicker conversion rate results in a greater total mass of planetesimals. The overall effect increases the total planetesimal mass, though not as significantly as the amplification of $\zeta$.

The radial resolution is a numerical factor that can influence simulation results. To assess its impact, we increased the number of refined cells in the infall region from 60 to 80 and 100 in the fiducial simulation. The total mass of planetesimals formed increased slightly from 9.97 $M_{\oplus}$ to 10.80 $M_{\oplus}$ and 11.32 $M_{\oplus}$, respectively. 
Although the results are not fully convergent with respect to radial resolution, this minor variation does not impact our conclusions.

\subsection{Caveats}

\subsubsection{Rossby wave instability}

The Rossby wave instability (RWI) can be triggered in a narrow, axisymmetric pressure bump by strong pressure gradients and non-Keplerian radial shear \citep{1999ApJ...513..805L,2000ApJ...533.1023L}. RWI generates anticyclonic vortices, which can disrupt the axisymmetric ring-like pressure bump. However, these vortices create azimuthal pressure bumps that efficiently trap dust \citep{2012A&A...545A.134M}, where the dust-to-gas ratio may exceed that of an axisymmetric pressure bump, potentially triggering streaming instability. 

To assess the stability of the infall-induced pressure bumps in this study against RWI, we applied the criterion for RWI proposed by \citet{2023ApJ...946L...1C},
\begin{equation}
    \kappa^2 + N_r^2 < 0.6 \Omega_{\rm K}^2,
\end{equation}
where $\kappa$ is the epicyclic frequency, and $N_r$ is the radial buoyancy frequency. 

In our fiducial simulation, an RWI-unstable region forms between 40.5 and 42.9 au, persisting from 14 to 51 kyr. Assuming no diffusion (calculating $\Sigma_{\rm g}$ using Eq. (\ref{eq:no_diff})), 1788 of the 4096 Sobol-sampled simulations remain stable against RWI, and 215 of the 653 simulations where planetesimal formation occurs are stable. This is a conservative estimate, as diffusion reduces the magnitude of the pressure gradient, lowering the likelihood of RWI.

Simulations have demonstrated that RWI vortices driven by infall are possible \citep{2015ApJ...805...15B,2022ApJ...928...92K}. However, the impact of infall-induced RWI on planetesimal formation --- whether positive or negative --- remains uncertain and requires further investigation through 2D or 3D simulations.

\subsubsection{Planetary evolution}

In our simulations, the planetesimal surface density profile and total planetesimal mass cease evolving once the conditions for streaming instability are no longer satisfied. However, a planetesimal belt with a width of only a few au and a total mass of just a few Earth masses is unlikely the ultimate fate of the system, as planetary evolution and its feedback on the protoplanetary disk have not been considered.

Since the planetesimals are formed in a dust-rich environment, they are expected to grow efficiently through pebble accretion \citep{2010A&A...520A..43O,2012A&A...544A..32L}. Pebble accretion might lead to a continuous increase in the total planetesimal mass even after the formation phase ends, accelerate the depletion of dust mass, and cause an earlier cessation of planetesimal formation due to rapid dust consumption.

Gas accretion begins when a planetary embryo reaches the pebble isolation mass \citep{2014A&A...572A..35L}, transforming a fraction of the gas in the disk into planetary mass. A sufficiently massive planet can carve a gap in the disk \citep{2017PASJ...69...97K,2020ApJ...889...16D}, creating a new pressure bump outside the gap. This might lead to the formation of a new generation of planetesimals, resulting in sequential planetesimal formation, as explored by \citet{2024A&A...688A..22L}.

The aerodynamic drag \citep{1976PThPh..56.1756A} and planet-disk interaction torques \citep{1979ApJ...233..857G,2002ApJ...565.1257T} generally cause planets to lose angular momentum and migrate inward. The pressure bump induced by infall might serve as a migration trap eliminating such torques and retaining planets \citep{2016MNRAS.460.2779C,2020A&A...638A...1M}. The growth of planets and their interactions with the disk following planetesimal formation in an infall-induced pressure bump merit further investigation through N-body simulations.

\section{Conclusions} \label{sec:conc}

This work demonstrates the feasibility of planetesimal formation from dust in a pressure bump created by the late-stage infall of interstellar medium onto a protoplanetary disk. We used \texttt{DustPy} to simulate gas evolution, dust transport and growth, and the transformation of dust into planetesimals triggered by streaming instability.

The Gaussian-distributed infalling gas generates a local maximum in the radial profile of gas surface density, leading to a local maximum in midplane pressure slightly inward of the gas surface density peak. This pressure bump halts the inward drifting dust from the outer disk and traps the infalling dust, enhancing the local dust-to-gas ratio and accelerating dust growth. When the critical conditions for streaming instability, governed by the midplane dust-to-gas ratio, are satisfied, planetesimals begin to form continuously, converting the dust surface density into planetesimal surface density. As the pressure bump is gradually smoothed out by viscous diffusion of gas, planetesimal formation ceases when the dust trap vanishes and the peak midplane dust-to-gas ratio drops below the threshold for streaming instability. Ultimately, the viscous spreading of the bump results in the outside-in formation of a planetesimal belt inside the central infall location.

A large number of simulations were conducted to investigate the effects of various parameters related to disk and infall properties. These parameters influence whether planetesimal formation occurs, the timing of formation, and the total mass of planetesimals produced. Planetesimal formation is favored by conditions where a massive, narrowly distributed infall of interstellar medium occurs onto a disk with low viscosity, low mass, extended surface density distribution, and high dust fragmentation velocity. Formation is highly unlikely when $\alpha > 2.5\times10^{-4}$ due to rapid gas diffusion, or when $v_{\rm frag} < 200\,\rm cm\,s^{-1}$ due to the fragility of dust particles. The critical conditions for planetesimal formation within an infall-induced pressure bump can be characterized by specific timescale relationships: $t_{\rm visc} \gtrsim 53.55\, t_{\rm drift}$, $t_{\rm visc} \gtrsim 10\, t_{\rm grow}$. These timescales are directly determined by the system's parameter setup. These quantitative conditions encapsulate the essential qualitative requirements for planetesimal formation: the creation of a prominent pressure bump that efficiently traps dust and its prolonged persistence. The primary source of material for planetesimals is the infalling dust; however, under favorable conditions, planetesimal formation can still occur even if the infalling material is solely gas.

In this study, we focused on the conditions and processes leading to planetesimal formation, but the subsequent stages of planetary evolution remain unexplored. Future work will extend our model to include the growth of planets through pebble and gas accretion, the dynamical interactions between planets, and their interactions with the gaseous disk. To achieve this, we need to couple our current dust and gas evolution code with an $N$-body simulation framework, allowing us to comprehensively model the evolution of a planetary system starting from the initial formation of planetesimals within an infall-induced pressure bump.

\begin{acknowledgements}
We acknowledge funding from the European Union under the European Union's Horizon Europe Research and Innovation Programme 101124282 (EARLYBIRD) and 101040037 (PLANETOIDS) and funding by the Deutsche Forschungsgemeinschaft (DFG, German Research Foundation) under grant 325594231, and Germany's Excellence Strategy - EXC-2094 - 390783311. Views and opinions expressed are, however, those of the authors only and do not necessarily reflect those of the European Union or the European Research Council. Neither the European Union nor the granting authority can be held responsible for them.
\end{acknowledgements}

\bibliographystyle{aa}

\end{document}